\documentclass[a4paper,11pt]{article}
\pdfoutput=1 

\usepackage{jheppub} 

\usepackage[T1]{fontenc} 
\usepackage{siunitx}

\usepackage{lineno}
\usepackage[perpage]{footmisc}

\title{\boldmath Identifying quenched jets in heavy ion collisions with machine learning}

\author[a]{Lihan Liu,}
\author[a]{Julia Velkovska,}
\author[b]{Marta Verweij}

\affiliation[a]{Department of Physics and Astronomy\\ 
 Vanderbilt University\\
 PMB 401807, 2301 Vanderbilt Place, Nashville, TN 37235, USA}
\affiliation[b]{Department of Physics\\ 
 Utrecht University\\
 Heidelberglaan 8 3584 CS Utrecht, The Netherlands}

\emailAdd{lihan.liu@vanderbilt.edu}
\emailAdd{julia.velkovska@vanderbilt.edu}
\emailAdd{m.verweij@uu.nl}

\abstract{Measurements of jet substructure in ultra-relativistic heavy ion collisions suggest that the jet showering process is modified by the interaction with the quark--gluon plasma. Modifications of the hard substructure of jets can be explored with modern data-driven techniques. In this study, a machine learning approach to the identification of quenched jets is designed. Jet showering processes are simulated with a jet quenching model \textsc{Jewel} and a non-quenching model \textsc{Pythia 8}. Sequential substructure variables are extracted from the jet clustering history following an angular-ordered sequence and are used in the training of a neural network built on top of a long short-term memory network. We show that this approach successfully identifies the quenching effect in the presence of the large uncorrelated background of soft particles created in heavy--ion collisions.}

\begin{document} 
\maketitle
\flushbottom


\section{Introduction}
\label{sec:intro}

Highly energetic partons (quarks and gluons) are expected to lose energy while traversing the extremely hot and dense quark-gluon plasma (QGP) created in ultra-relativistic heavy ion collisions~\cite{Bjorken:1982tu}. This phenomenon, known as ``jet quenching'', was discovered at the Relativistic Heavy Ion Collider (RHIC) through the observation of suppression of high transverse momentum ($p_{\mathrm{T}}$) hadrons~\cite{Adcox:2001jp,STAR:2002ggv,Adler:2003qi,Adams:2003kv} and back-to-back dihadron correlations~\cite{STAR:2002svs}. These observations were confirmed at the Large Hadron Collider (LHC) and extended into a larger kinematic range~\cite{Aamodt:2010jd,Aamodt:2011vg,CMS:2012aa,Aad:2015wga, Khachatryan:2016odn}.  

In addition to measurements on high $p_\mathrm{T}$ hadrons, which are the leading fragments of jets, the LHC opened new opportunities for direct observations of jet quenching and jet modifications in the QGP. Significant dijet transverse momentum asymmetry~\cite{ATLAS:2010isq, CMS:2011iwn,Chatrchyan:2012nia,Khachatryan:2015lha} and suppressed jet production~\cite{Aad:2012vca,Abelev:2013kqa,Adam:2015ewa} were observed in lead-lead (PbPb) collisions. Jet-hadron correlations gave insights into how the energy lost by the jets is distributed in 
the medium produced in the collisions. Measurements of the jet shapes (transverse momentum radial profile)~\cite{Chatrchyan:2013kwa,ALICE:2019whv,ATLAS:2019pid} and the jet fragmentation functions~\cite{Chatrchyan:2014ava,Aaboud:2017bzv} were performed to further investigate the mechanism of the jet shower modifications in the QGP. To quantify the medium modifications, measurements in proton-proton (pp) collisions~\cite{CMS:2021iwu,CMS:2018vzn,ATLAS:2019mgf,ATLAS:2019kwg,ATLAS:2020bbn,ATLAS:2017zda,ALICE:2021njq,ALICE:2022hyz,ALICE:2017nij} are used as a reference for jet modifications in the medium.
The jet splitting function has been measured in both pp and PbPb collisions~\cite{Sirunyan:2017bsd, STAR:2021kjt,ALargeIonColliderExperiment:2021mqf}, through the usage of a jet grooming algorithm that is able to split (``decluster'') a single jet into two subjets and locate the hard splitting by removing softer wide-angle radiation contributions. The hard splitting, characterized by two well-separated subjets (``two-prong'' structure), provides access to the early stages in the parton shower evolution. 

In recent years many machine learning applications have been developed for studies in high-energy physics \cite{Feickert:2021ajf}. For the identification of quenched jets, a previous study of this classification problem~\cite{Apolinario:2021olp} explored several ways of jet representation and different neural network architectures. It showed some promising results of deep learning techniques applied for the study of quenched jets in QGP at simulation level. However, the application of the convolutional neural network (CNN) often involves image preprocessing techniques, such as rotation, and normalization of the pixel intensity, which are not universal. These techniques are usually designed for a specific jet species, for example the boosted W/Z jets as studied in Ref.~\cite{Macaluso:2018tck}, and may use a particular jet substructure feature, as studied in Ref.~\cite{Du:2020pmp,Du:2021pqa} for QCD jets. In addition, as a common practice in the implementation of modern Monte Carlo event generators, events associated with small cross-sections which often produce high $p_\mathrm{T}$ jets, can be over-sampled for statistical benefits, and event weights are assigned accordingly for compensation. How the event weights fit into the development of a machine learning approach is not well studied. Such problem is addressed in Ref.~\cite{Du:2020pmp} and a re-weighting method is designed. The sample weights are assigned based on the so-called effective sample number \cite{DBLP:journals/corr/abs-1901-05555}, which are not related to event cross-sections. For a more straight forward application to experimental data, it is desirable to have sample distributions corresponding to realistic cross-sections. Another challenge is that in the experiment the jets exist in the presence of a thermal background of uncorrelated particles created in heavy--ion collisions. Therefore, before the feasibility of machine learning can be established, techniques such as event mixing and background subtraction should be applied to simulated events, after which the jet finding algorithms can be performed. Such problem is addressed in Ref.~\cite{Du:2020pmp} and the authors found a decrease in the performance due to the presence of the underlying event. A similar conclusion is reached in  Ref.~\cite{Lai:2021ckt}.

In this study, we use the successive splittings of the parton shower as input for a long short-term memory (LSTM)~\cite{hochreiter1997long,Sherstinsky_2020} neural network. These splittings are obtained using jet declustering procedures in which all the splittings from the primary Lund plane~\cite{Dreyer:2018nbf} are extracted. The LSTM neural network is a special type of recurrent neural network that is capable of learning long-term dependencies and, therefore, has the potential to provide insight on the parton shower evolution in the medium. It has been used for jet tagging based on the primary Lund sequence~\cite{Dreyer:2018nbf,Dreyer:2020brq}, showing a reasonable performance in W tagging, Top tagging and quark/gluon discrimination. Such tagging problems have also been studied in heavy-ion collisions with CNN \cite{Chien:2018dfn}. In addition, the LSTM neural network shows its ability of jet grooming with a carefully designed training strategy~\cite{Carrazza:2019efs}. We use the LSTM neural network and perform supervised machine learning in which a well-trained classifier is able to distinguish quenched jets from vacuum jets on a jet-by-jet basis.

We present a machine learning approach to the identification of quenched jets in the presence of a large uncorrelated underlying event as present in heavy ion collisions. In Section~\ref{sec:event_simulation}, we describe the procedure of event simulation and the subtraction procedure on the uncorrelated underlying event. In Section~\ref{sec:machine_learning}, we describe how the supervised machine learning is performed, including a general feature engineering method on jets, a training method with event weights taken into consideration, a method of hyper-tuning and a study of robustness. In Section~\ref{sec:results}, we test the discrimination power of the trained classifier for quenched against non-quenched jets. Finally, we present our conclusions and outlook in Sec.~\ref{sec:conclusion}. The full code and data used in this study are available as open-source~\cite{VandyMLJets}.


\section{Data Sample}
\label{sec:event_simulation}

In this study, we simulate hard scattering events with Monte Carlo event generators \textsc{Pythia 8}~\cite{Sjostrand:2007gs} and \textsc{Jewel v2.2.0}~\cite{Zapp:2012ak,KunnawalkamElayavalli:2017hxo} at a center-of-mass energy per nucleon pair ($\sqrt{s_{\mathrm{NN}}}$) of $\SI{5.02}{TeV}$. A minimum outgoing parton transverse momentum $\hat p_{\mathrm{T}} = \SI{100}{GeV}$. \textsc{Pythia 8} is used to simulate dijet events in pp collisions, while \textsc{Jewel} simulates dijet events in heavy--ion collisions including parton-medium interactions. An important consequence of the parton-medium interactions implemented in \textsc{Jewel} is the medium response. In our studies, we use \textsc{Jewel} with keeping track of the recoils to explore the possible extreme. 
It has to be pointed out that such implementation of the medium response is a model-dependent component, while other generators may implement their own descriptions of the medium response. More details can be found in Ref.~\cite{Andrews:2018jcm}. 
To simulate the thermal background originating from the hadronization of all partons in the quark-gluon plasma, which is not correlated with the jets, dijet events from both event generators are embedded into underlying events, which have general features similar to those observed experimentally. Following the implementation in Ref.~\cite{Andrews:2018jcm}, the particles in the uncorrelated underlying events are generated from a Boltzmann distribution in $p_\mathrm{T}$, and uniform distribution in pseudorapidity $\eta$ and azimuthal angle $\varphi$. We prepare the uncorrelated underlying events using two different multiplicity settings, corresponding roughly to the mid-central (40-50\%) and the most-central (0-10\%) PbPb collisions at $\sqrt{s_{\mathrm{NN}}}=\SI{5.02}{TeV}$, as inferred from published experiment results~\cite{ALICE:2016fbt}. More details about the uncorrelated underlying events are listed in Table~\ref{tab:tab1}. With the inclusion of correlated and uncorrelated underlying event backgrounds, the influence on substructure variables introduced by particles not originating from the parton shower can be investigated. 

\begin{table}[h]
\begin{center}
\begin{tabular}{ c|c|c } 
 Values & Mid-central & Most-central \\ 
 \hline
 Pseudorapidity Interval & $|\eta|<3$ & $|\eta|<3$ \\
 Total Event Multiplicity & 1700 & 7000\\
 Corresponding Centrality Interval  & 40-50\% & 0-10\%\\ 
 Average Yield $\langle \frac{\mathrm{d}N}{\mathrm{d}\eta} \rangle$ & 283 & 1167 \\
 Average Transverse Momentum $\langle p_{\mathrm{T}} \rangle$ & $\SI{0.9}{GeV}$ & $\SI{1.2}{GeV}$ \\
 Average Transverse Momentum Density $\langle \rho \rangle$ & $\SI {35.0} {GeV/Area}$ & $\SI {227.1}{GeV/Area}$ \\
 \hline
\end{tabular}
\bigskip
\caption{Parameters of the simulated underlying events.}
\label{tab:tab1}
\end{center}
\end{table}

The mixed event, constructed by embedding a dijet event into an uncorrelated underlying event, contains a hard scattering in the presence of the soft thermal background in heavy--ion collisions. To eliminate the energy and momentum contribution of thermal background particles into the jets, we perform background subtraction on the mixed events. The event-wide constituent subtraction method~\cite{Berta:2014eza} is used with maximum distance parameter between the signal and background particles $\Delta R_{\mathrm{max}} = 0.3$, using $p_{\mathrm{T}}$ weight $\alpha=1$. Particles that remain after subtraction are clustered into jets with the anti-$k_\mathrm{T}$ algorithm~\cite{Cacciari:2008gp} with a distance parameter $R=0.4$.

\begin{figure}[ht]
    \centering
    \includegraphics[width=0.328\textwidth]{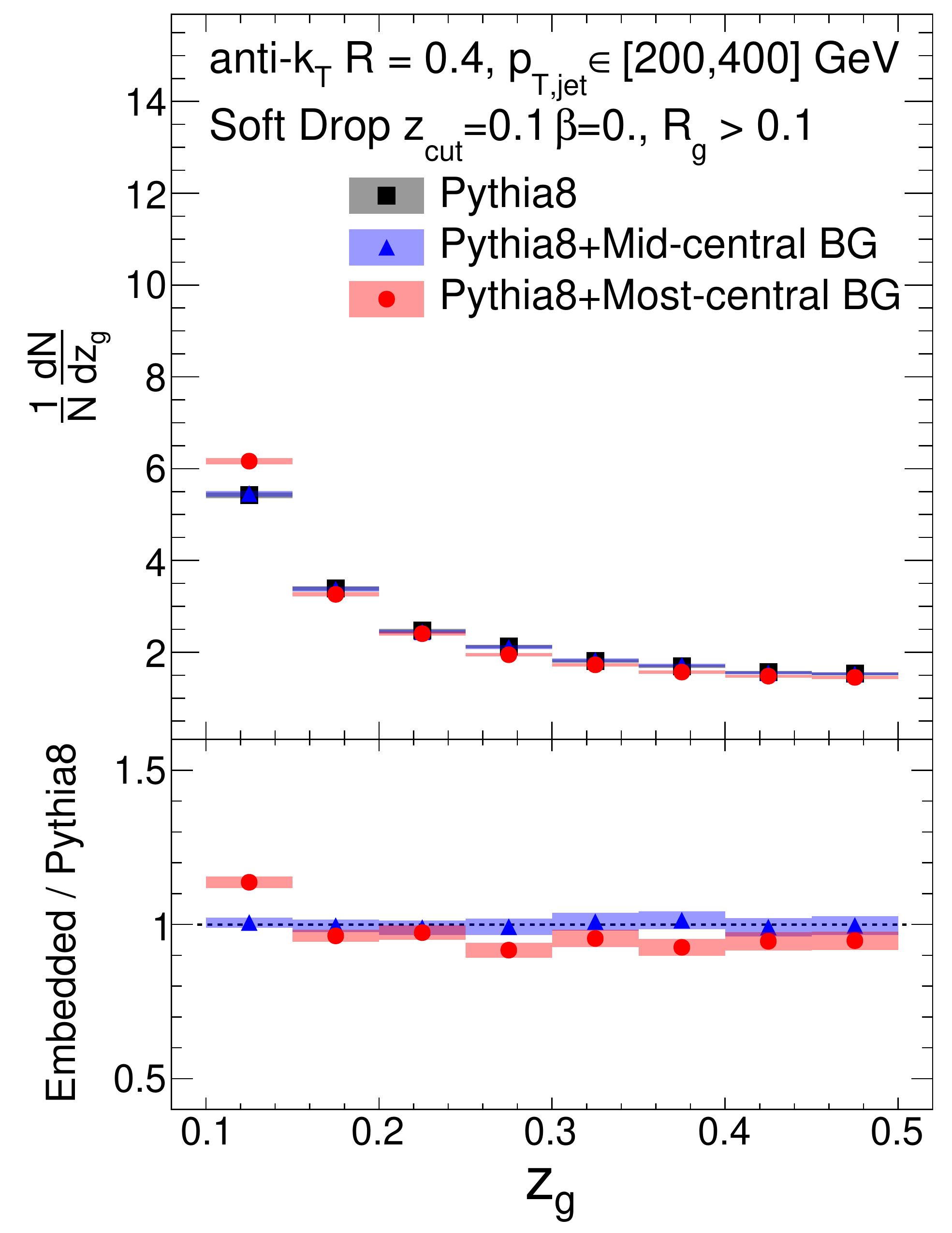}
    \includegraphics[width=0.328\textwidth]{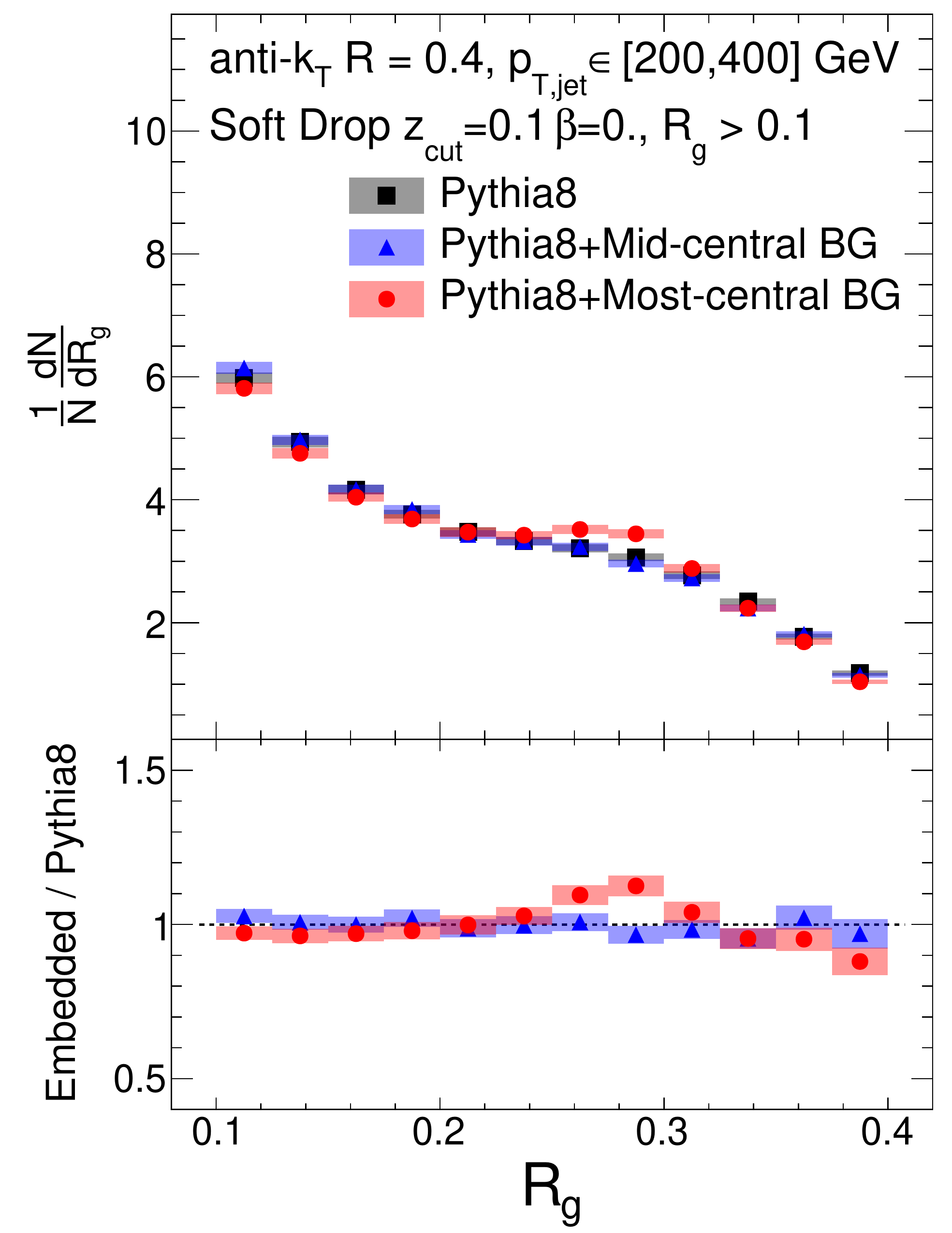}
    \includegraphics[width=0.328\textwidth]{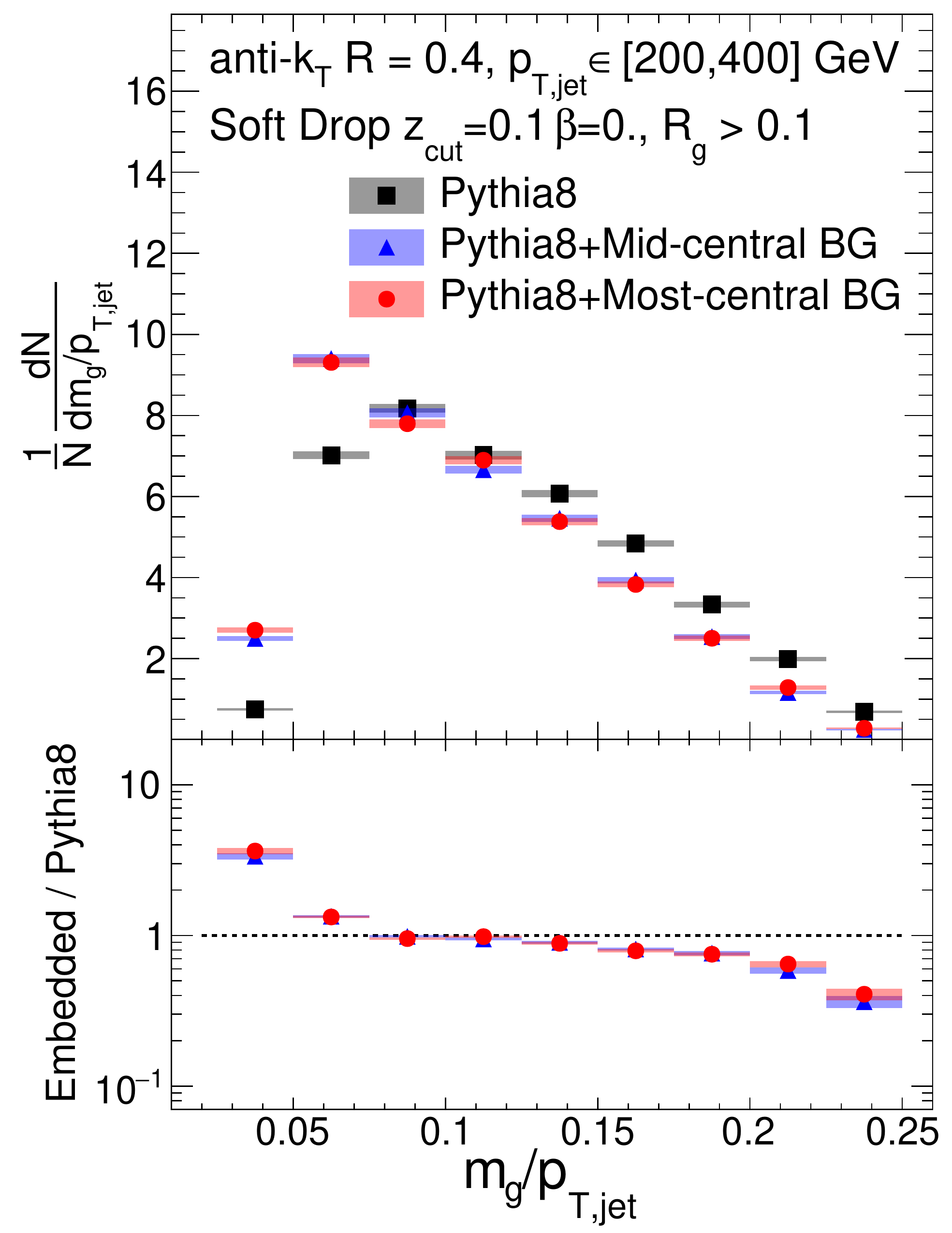}
    \caption{Distribution of substructure variables of the groomed jets with and without embedding.}
    \label{fig:sub}
\end{figure}

After jet finding, each jet is reclustered with the Cambridge/Aachen algorithm~\cite{Dokshitzer:1997in} from its constituents to form a pairwise and angular-ordered structure. We then apply the soft drop declustering~\cite{Larkoski:2014wba} procedure, continuously removing soft branches and keeping the splittings that satisfy the soft drop condition,
\begin{equation}
    z_{\mathrm{g}} \equiv \frac{\mathrm{min}(p_{\mathrm{T}1} , p_{\mathrm{T}2})}{p_{\mathrm{T}1}+p_{\mathrm{T}2}} > z_{\mathrm{cut}} \left( \frac{\Delta R}{R_0} \right) ^\beta,
    \label{eq:sd}
\end{equation}
where $p_{\mathrm{T}1}$ and $p_{\mathrm{T}2}$ are the transverse momenta of the two separated subjets, $\Delta R$ is their angular distance, and $R_0$ is the jet distance parameter, in our case $R_0 = 0.4$. In this study, we set the free parameters to values $z_{\mathrm{cut}} = 0.1$, and $\beta = 0$. We also applied a cut of $p_{\mathrm{T,jet}}> \SI{200}{GeV}$ to optimize the purity in locating the hard splittings~\cite{Mulligan:2020tim}. To evaluate the background subtraction procedure in Fig.~\ref{fig:sub} we compare the distributions of jet  substructure variables for jets simulated with \textsc{Pythia} and for \textsc{Pythia} jets embedded into thermal background events  with particle multiplicities corresponding to mid-central and most-central PbPb collision events after the background subtraction. Ideally, if the background subtraction corrects both the jet energy and the jet substructure, the results will be identical. The top panels of Fig.~\ref{fig:sub} show the jet splitting function $z_{\mathrm{g}}$, the angular distance between the subjets $R_{\mathrm{g}}$, and the groomed jet mass $m_{\mathrm{g}}$ divided by the jet $p_{\mathrm{T}}$, for jets propagating in vacuum (simulated with \textsc{Pythia}) before and after the embedding into the underlying PbPb events and the corresponding background subtraction. The bottom panels show the closure test of the subtraction procedure, e.g. the ratio of the jet substructure variables for jets embedded in the PbPb-backgrounds after subtraction, and the original \textsc{Pythia} jets. We find that for most of the phase space of interest the subtraction procedure works well, although there are some discrepancies that can be associated with mis-tagging of the subleading subjet due to fluctuations in the thermal background, as suggested in Ref.~\cite{Mulligan:2020tim}. This indicates that large background fluctuations may produce effects that mimic a jet quenching signal when the soft drop grooming is applied. Other grooming methods may help control the mis-tagging and could be a subject of further investigations. In this study, we use the soft drop grooming imposing a jet $p_{\mathrm{T}}$ cut of $p_{\mathrm{T,jet}}> \SI{200}{GeV}$, which reduces mis-tagging.

\begin{figure}[htb]
    \centering
    \includegraphics[width=0.328\textwidth]{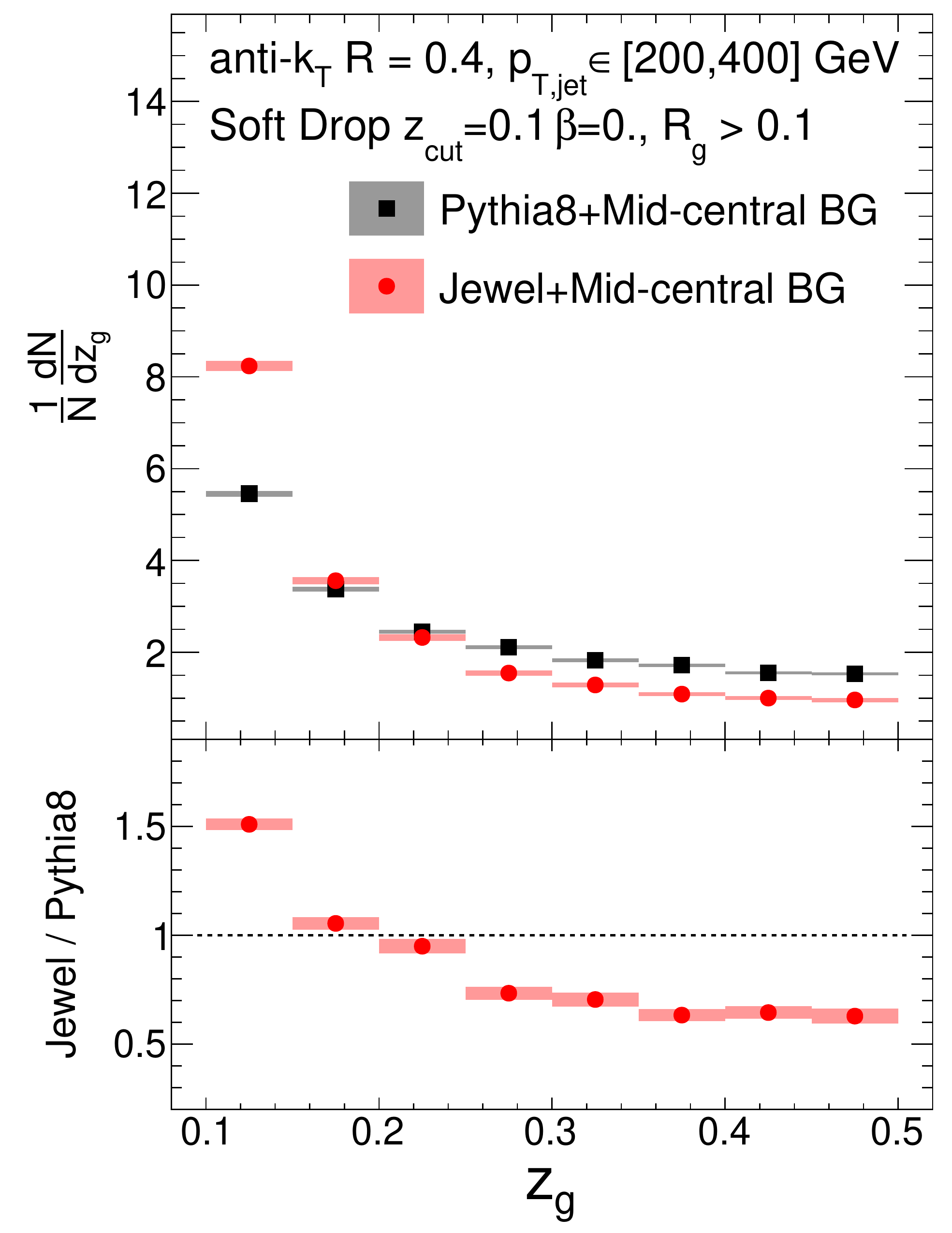}
    \includegraphics[width=0.328\textwidth]{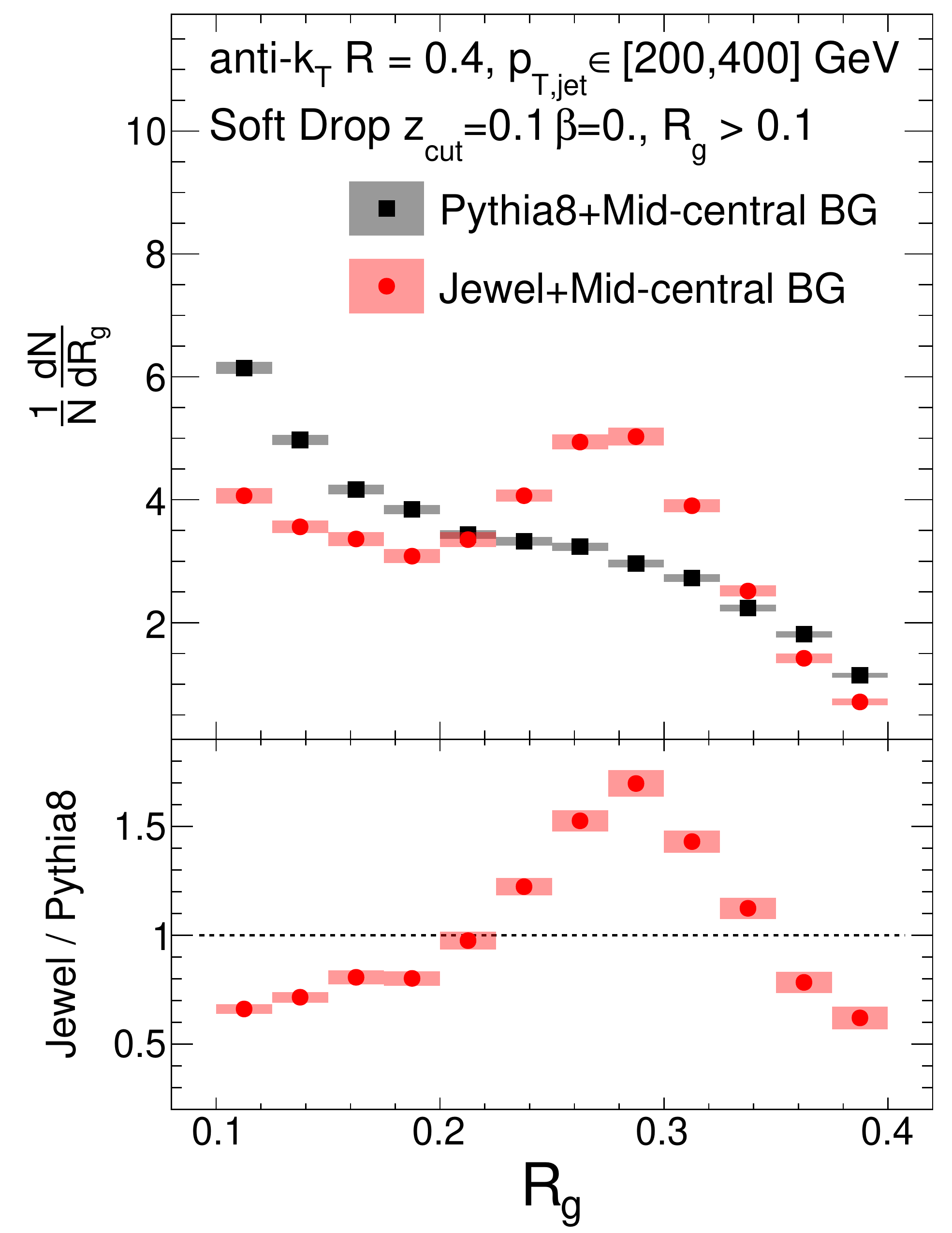}
    \includegraphics[width=0.328\textwidth]{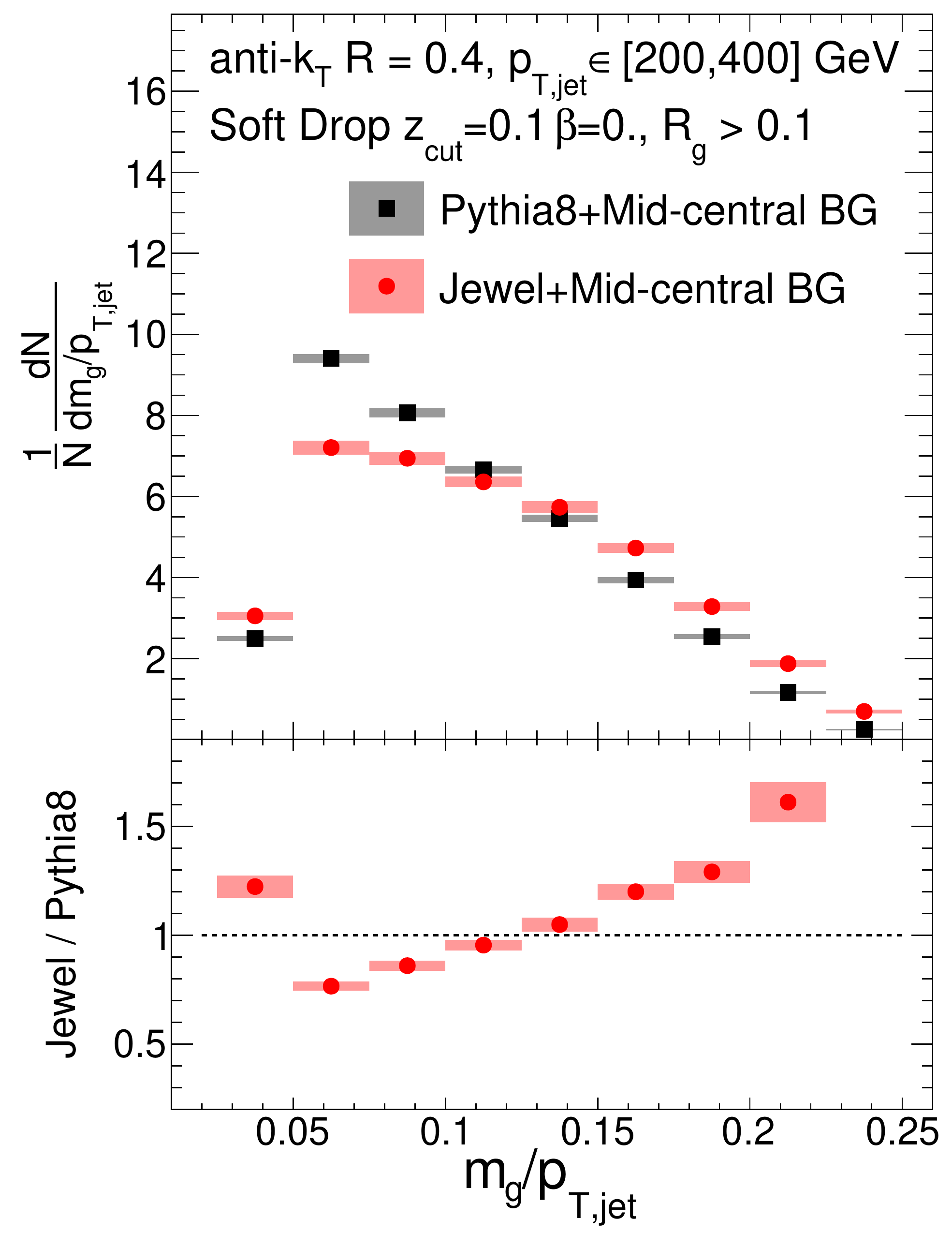}
    
    \vspace{0.1in}
    
    \includegraphics[width=0.328\textwidth]{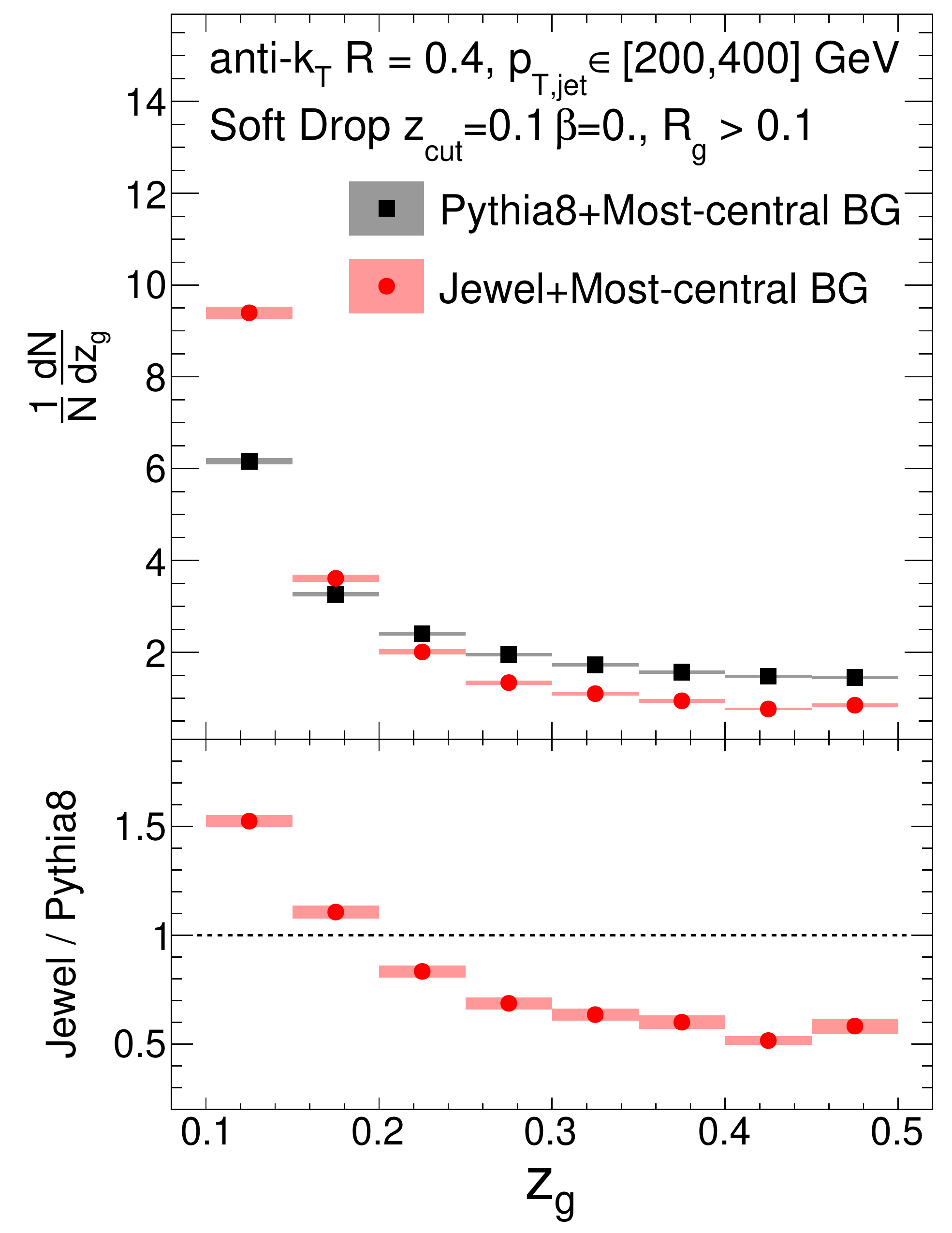}
    \includegraphics[width=0.328\textwidth]{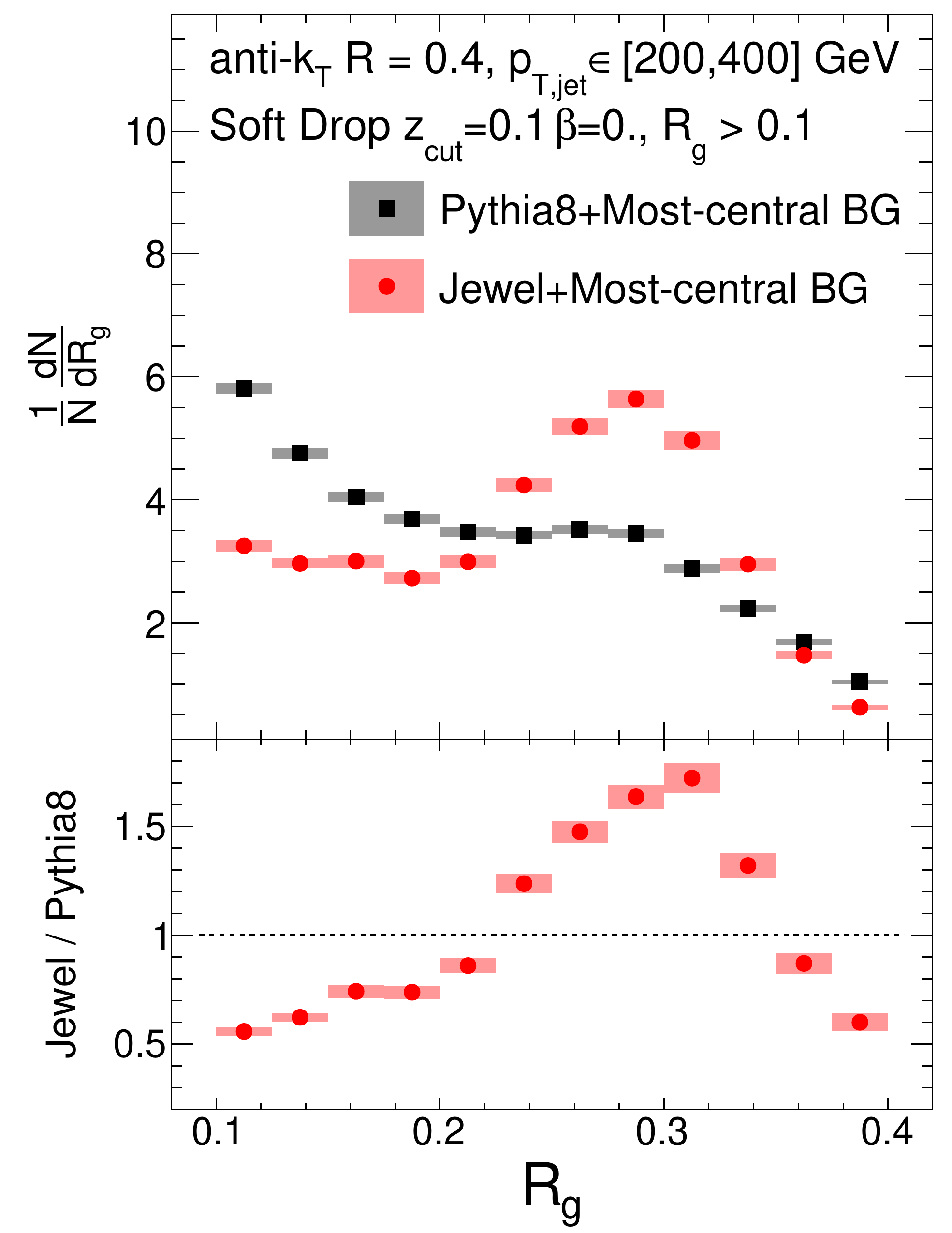}
    \includegraphics[width=0.328\textwidth]{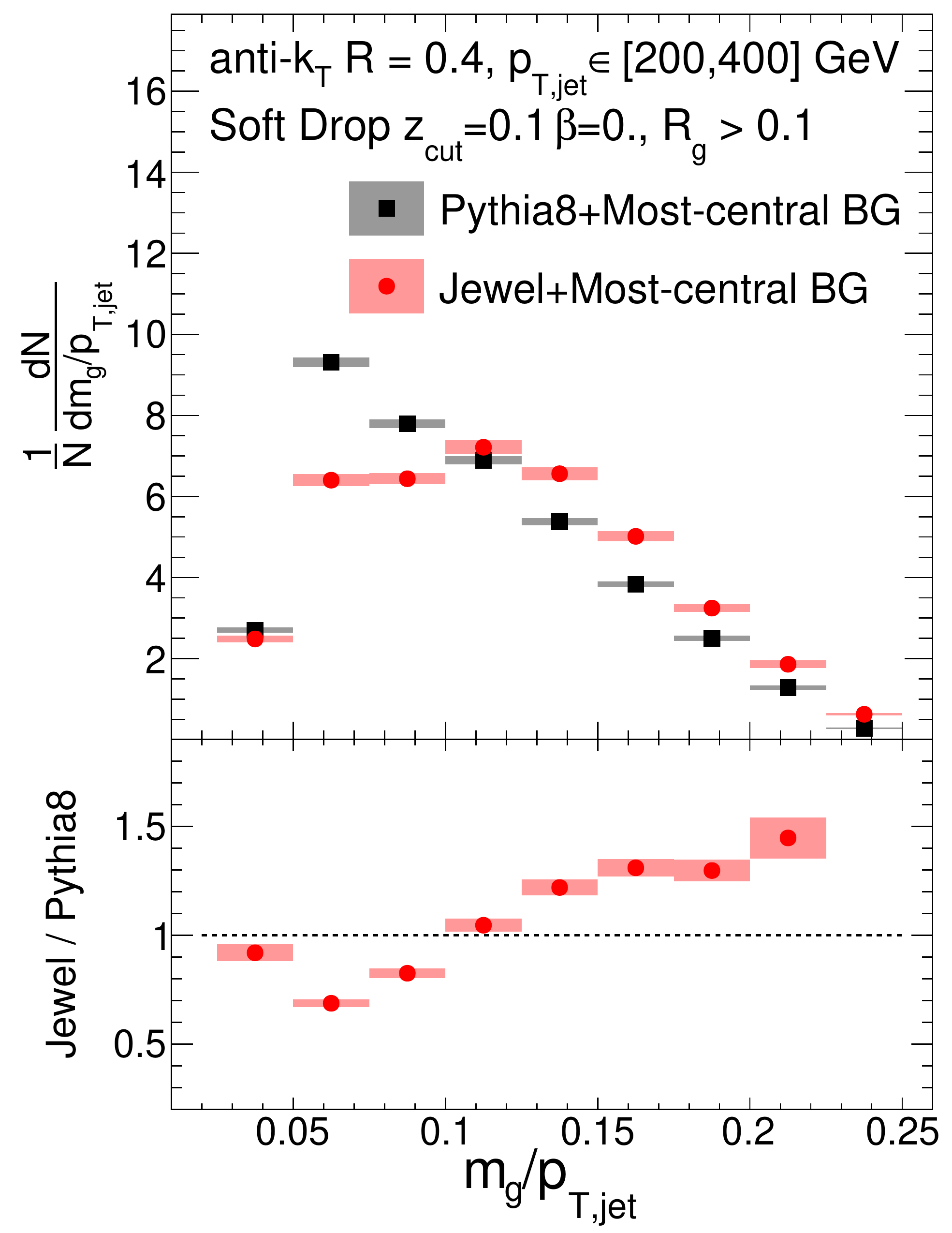}
    \caption{Distribution of substructure variables of the groomed jets for the mid-central (upper) and the most-central (lower) mixing scenarios.}
    \label{fig:comp}
\end{figure}

\begin{figure}[htb]
    \centering
    \includegraphics[width=0.48\textwidth]{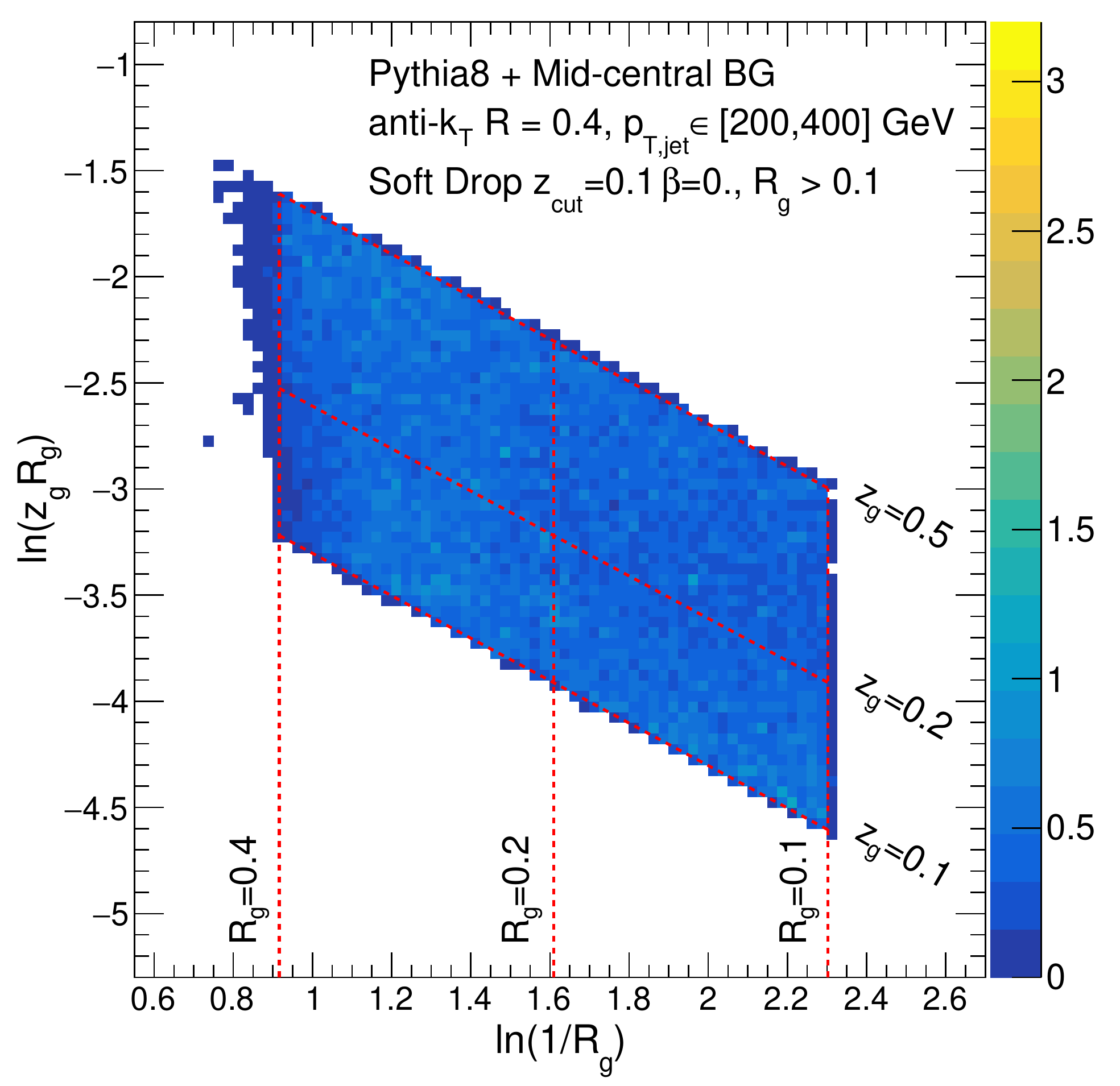}
    \includegraphics[width=0.48\textwidth]{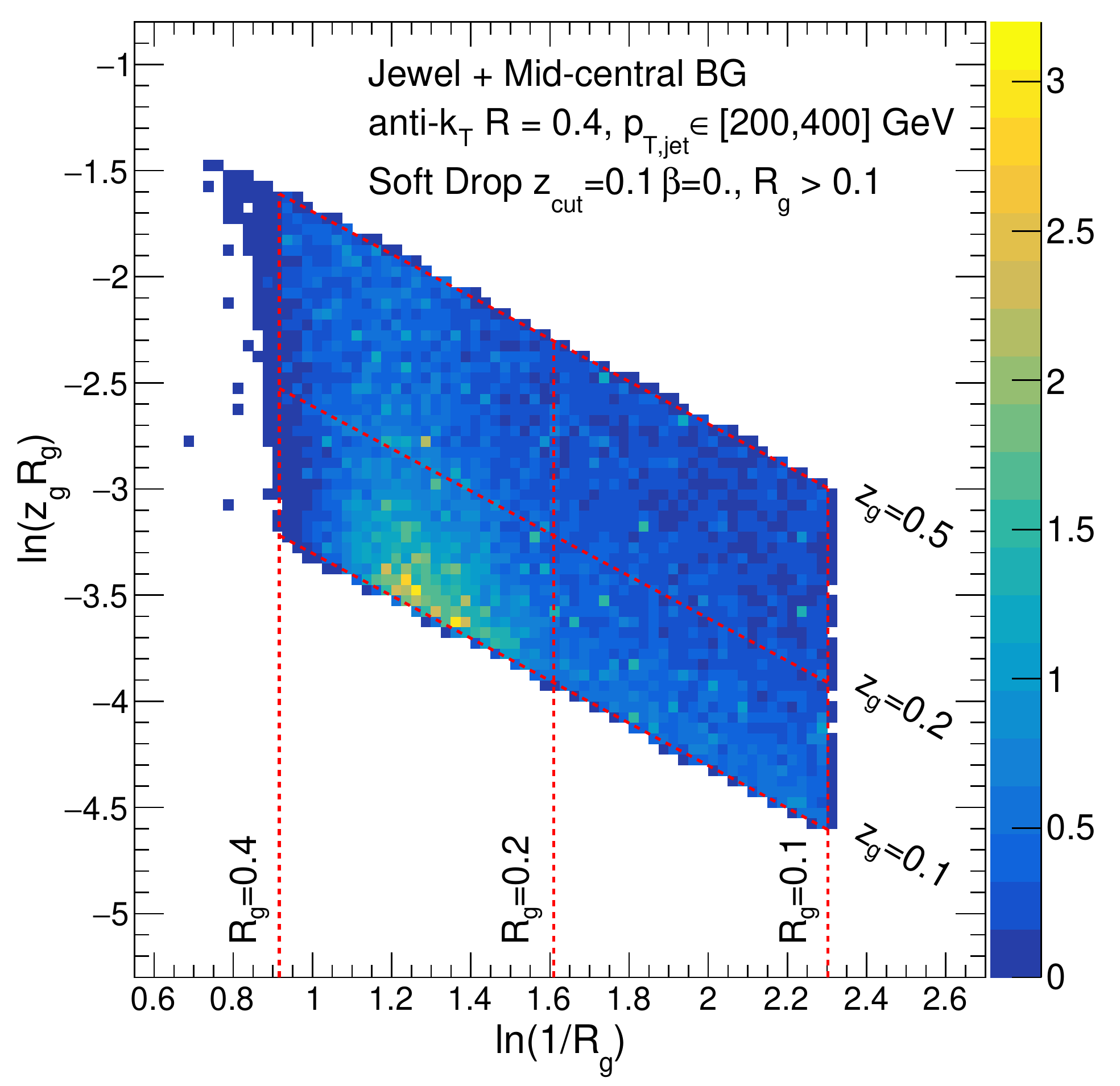}
    
    \vspace{0.1in}
    
    \includegraphics[width=0.48\textwidth]{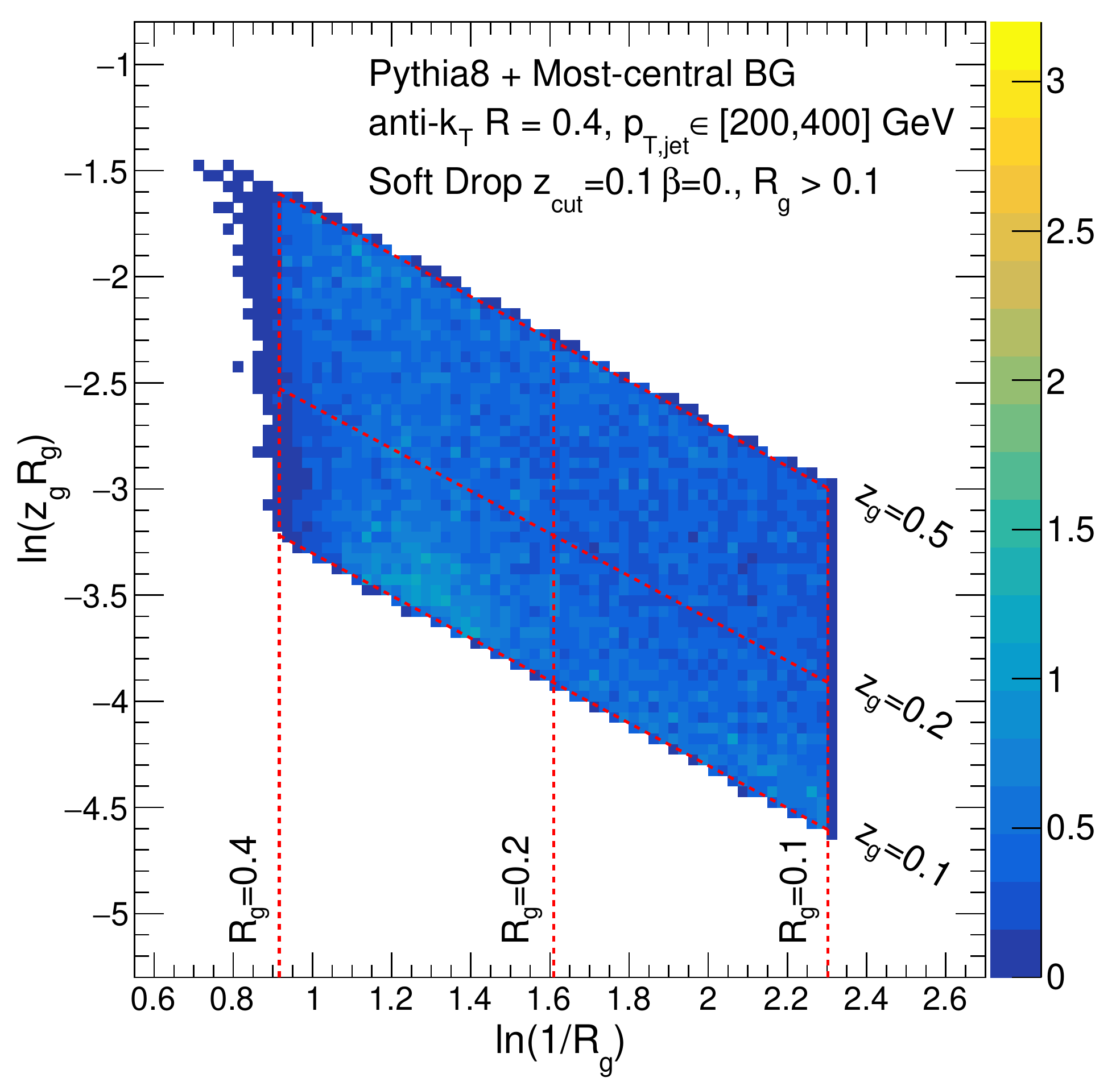}
    \includegraphics[width=0.48\textwidth]{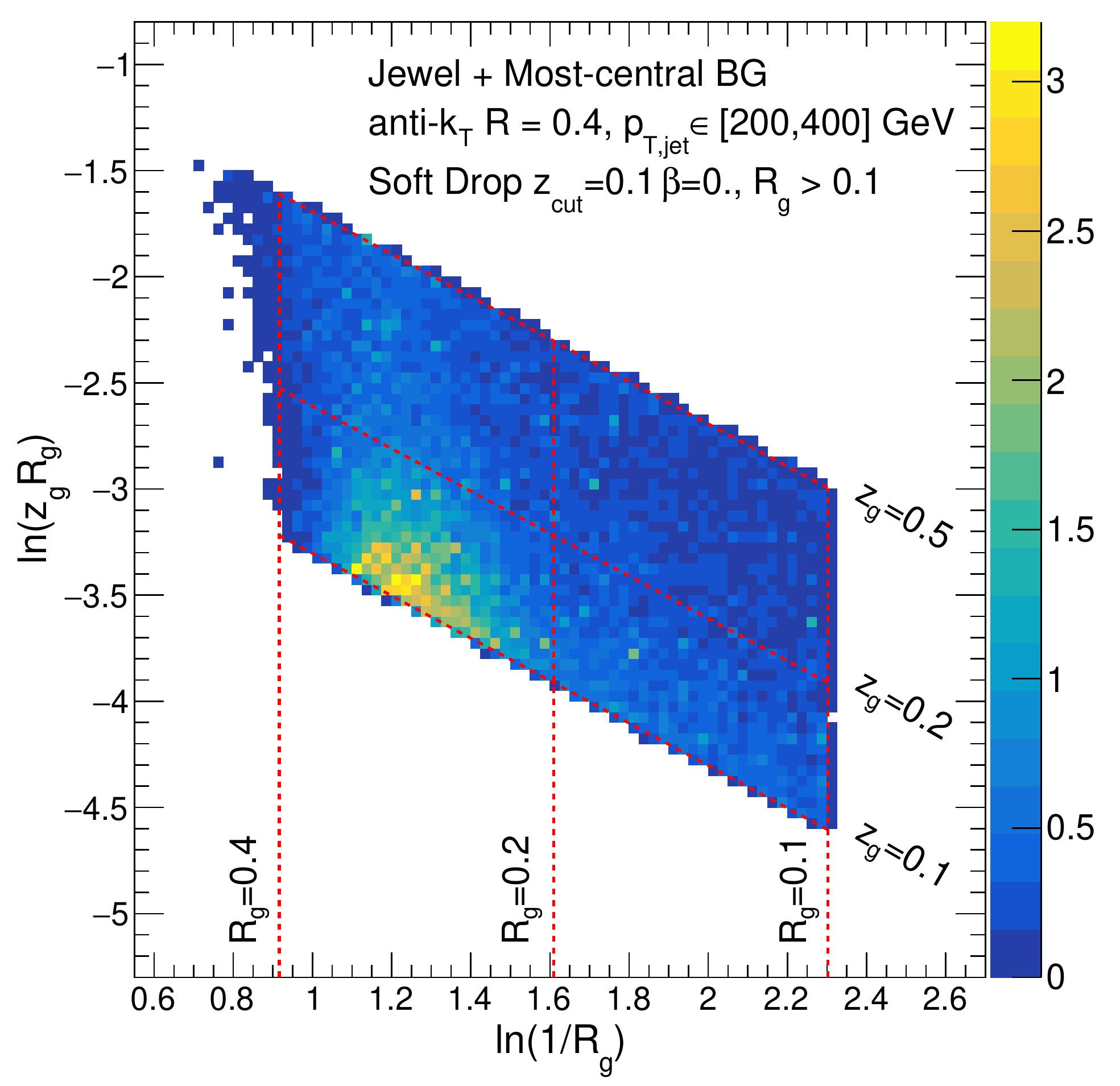}
    \caption{Primary Lund plane density for the mid-central (upper) and the most-central (lower) mixing scenarios. }
    \label{fig:lund}
\end{figure}

To investigate the quenching effects on groomed jets, we compare the substructure variables of quenched jets with respect to vacuum jets, where both undergo embedding and background subtraction. The distributions of substructure variables of the groomed jets in the mid-central and the most-central mixing scenarios are shown in Fig.~\ref{fig:comp}. Significant modifications are seen on all jet substructure variables. Additionally, in Fig.~\ref{fig:lund} we plot the primary Lund plane density~\cite{Dreyer:2018nbf},
\begin{equation}
\centering
\rho(z_{\mathrm{g}}R_{\mathrm{g}}, R_{\mathrm{g}}) = \frac{1}{N_{\mathrm{jet}}} \frac{\mathrm{d}^2N}{\mathrm{d}\,ln(z_{\mathrm{g}}R_{\mathrm{g}})\,\mathrm{d}\,ln(1/R_{\mathrm{g}})},
\end{equation}
where $z_{\mathrm{g}}$ and $R_{\mathrm{g}}$ are taken from the first pair of hard branches that pass the soft drop condition (Eq.~\ref{eq:sd}) for each jet. Strong modifications of the hard splittings with enhancement in wider (larger $R_{\mathrm{g}}$) and more unbalanced (lower $z_{\mathrm{g}}$) splittings are apparent in the \textsc{Jewel} events. Next, we implement a machine learning approach to classify quenched and non-quenched jets. 

\clearpage

\section{Supervised Machine Learning}
\label{sec:machine_learning}

\subsection{Feature Engineering}
\label{sec:feature_engineering}

The feature engineering procedure starts from jets reclustered with the Cambridge/Aachen algorithm. This leads to a clustering history with a binary tree structure, of which branches are separated based on their angular distance.
\begin{figure}[htp]
    \centering
    \includegraphics[width=0.8\textwidth]{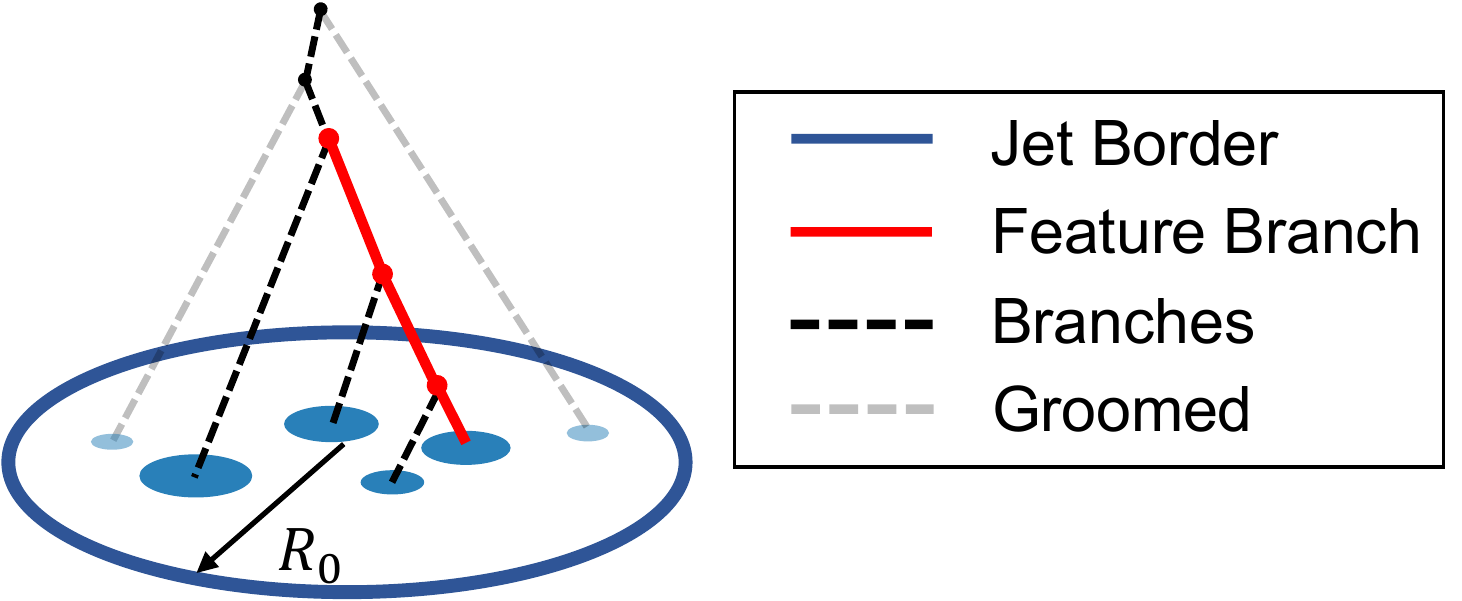}
    \caption{Jet clustering history with a binary tree structure.}
    \label{fig:bst}
\end{figure}
The soft drop groomer continuously removes the softer branch, until finding the hard splitting that satisfies the soft drop condition (Eq.~\ref{eq:sd}), after which the declustering continues on the harder subjet, as illustrated in Fig.~\ref{fig:bst}.  

Each splitting can be described with a feature vector, $x_{\mathrm{t}}$, consisting of substructure variables including, but not limited to, the momentum fraction $z$, angular distance $\Delta R$, perpendicular momentum $k_\perp$, and the invariant jet mass $m_{\mathrm{inv}}$, which are all calculated from the pair of declustered subjets with:
\begin{equation}
\begin{aligned}
\centering
     & z = \frac{\mathrm{min}(p_{\mathrm{T,i}} , p_{\mathrm{T,j}})}{p_{\mathrm{T,i}}+p_{\mathrm{T,j}}},\\
     & \Delta R = \sqrt{(\phi_{\mathrm{i}}-\phi_{\mathrm{j}})^2+(\eta_{\mathrm{i}}-\eta_{\mathrm{j}})^2},\\
     & k_{\perp} = \mathrm{min}(p_{\mathrm{T,i}} , p_{\mathrm{T,j}}) * \Delta R,\\
     & m_{\mathrm{inv}} = \sqrt{(E_{\mathrm{i}}+E_{\mathrm{j}})^2-(\textbf{p}_{\mathrm{i}}+\textbf{p}_{\mathrm{j}})^2},\\
     & x_{\mathrm{t}} = [z, \Delta R, k_{\perp}, m_{\mathrm{inv}}],\\
\end{aligned}
\end{equation}
where $i, j$ denote subjets at the declustering step $t$. In this way, sequential vectors $[x_0,...,x_{\mathrm{t}},...]$ are extracted, and are then used in the training of a neural network.


\subsection{Training of LSTM-based Neural Network}

A classification problem, that is, whether the hot and dense QGP medium modifies the jet showering process, is investigated based on a supervised machine learning strategy. Jets simulated with \textsc{Pythia 8} are labeled as 0, in the later context referring to the non-quenched or the negative class. Conversely, those simulated with \textsc{Jewel} are labeled as 1, referring to the quenched or the positive class. Data samples are split into training and validation sets in a 1:1 ratio. For each jet, sequential feature vectors are extracted from their clustering history, as described in Section~\ref{sec:feature_engineering}. Considering that the lengths of the sequential feature vectors may vary, all the sequences in the same batch have post-sequence zeros as pads, such that they have the same length after padding. The zero-padded batch of sequential feature vectors is fed to a neural network built with LSTM layers and fully connected (FC) layers, of which the implementation is provided by the PyTorch framework~\cite{NEURIPS2019_9015}. Details about the parameters of the neural network architecture are further discussed in Section~\ref{subsec:hypertuning}.

\begin{figure}[htp]
    \centering
    \includegraphics[width=0.6\textwidth]{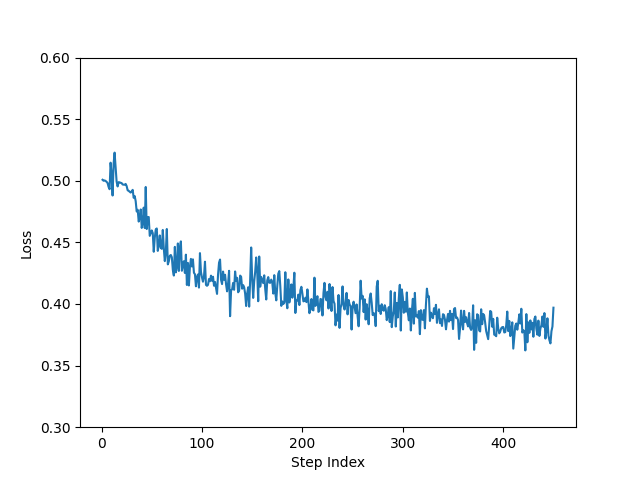}
    \caption{Examples of batch loss decreasing in the training process.} 
    \label{fig:loss}
\end{figure}

A common LSTM unit~\cite{Sherstinsky_2020} is composed of a cell, an input gate, an output gate, and a ``forget'' gate. The cell remembers relevant information throughout the processing of sequences, and the three gates regulate the flow of information into and out of the cell.
The output of the LSTM network at the last step\footnote{The output from LSTM layer has a dimension of [batch size, sequence length, hidden size] (\url{https://pytorch.org/docs/stable/generated/torch.nn.LSTM.html}). The hidden size is a configurable parameter, in other words a hyper-parameter.} is directed to FC layers which map high-dimensional values to a vector of predictive values, from which the batch losses are calculated. We use the weighted mean squared error (MSE) loss function:

\begin{equation}
\begin{aligned}
\centering
& l_{\mathrm{MSE}} = \frac{\sum\limits_{\mathrm{batch}} \omega_{\mathrm{i}} * (x_{\mathrm{i}} - y_{\mathrm{i}})^2}{\sum\limits_{\mathrm{batch}} \omega_{\mathrm{i}}},
\end{aligned}
\end{equation}
where $x_{\mathrm{i}},y_{\mathrm{i}}$ refer to the predictive label and the truth label of the $ith$ jet, and $\omega_{\mathrm{i}}$ is the event weight scaled from the event cross-section. During the training, learnable neural network parameters (e.g., coefficients or biases) are updated to minimize the batch loss during training, as shown in Fig.~\ref{fig:loss}, using the ``gradient decent'' algorithm. 

As mentioned previously, event weights come from generators due to over-sampling on small cross-section events, for the purpose of compensation. This is the default behavior in \textsc{Jewel}, but is optional in \textsc{Pythia8}. In our study the event weight is accounted for  in the batch loss calculation. It can easily be replaced with  unity for unbiased experimental samples. In the current study the introduction of event weights has brought challenges to the training process, showing significant loss fluctuation when batch size is small. Larger batch sizes help to regulate the batch loss fluctuation, with the trade-off of making the computation heavier.

\subsection{Hyper-tuning}
\label{subsec:hypertuning}
\begin{figure}[htp]
    \centering
    \includegraphics[width=0.49\textwidth]{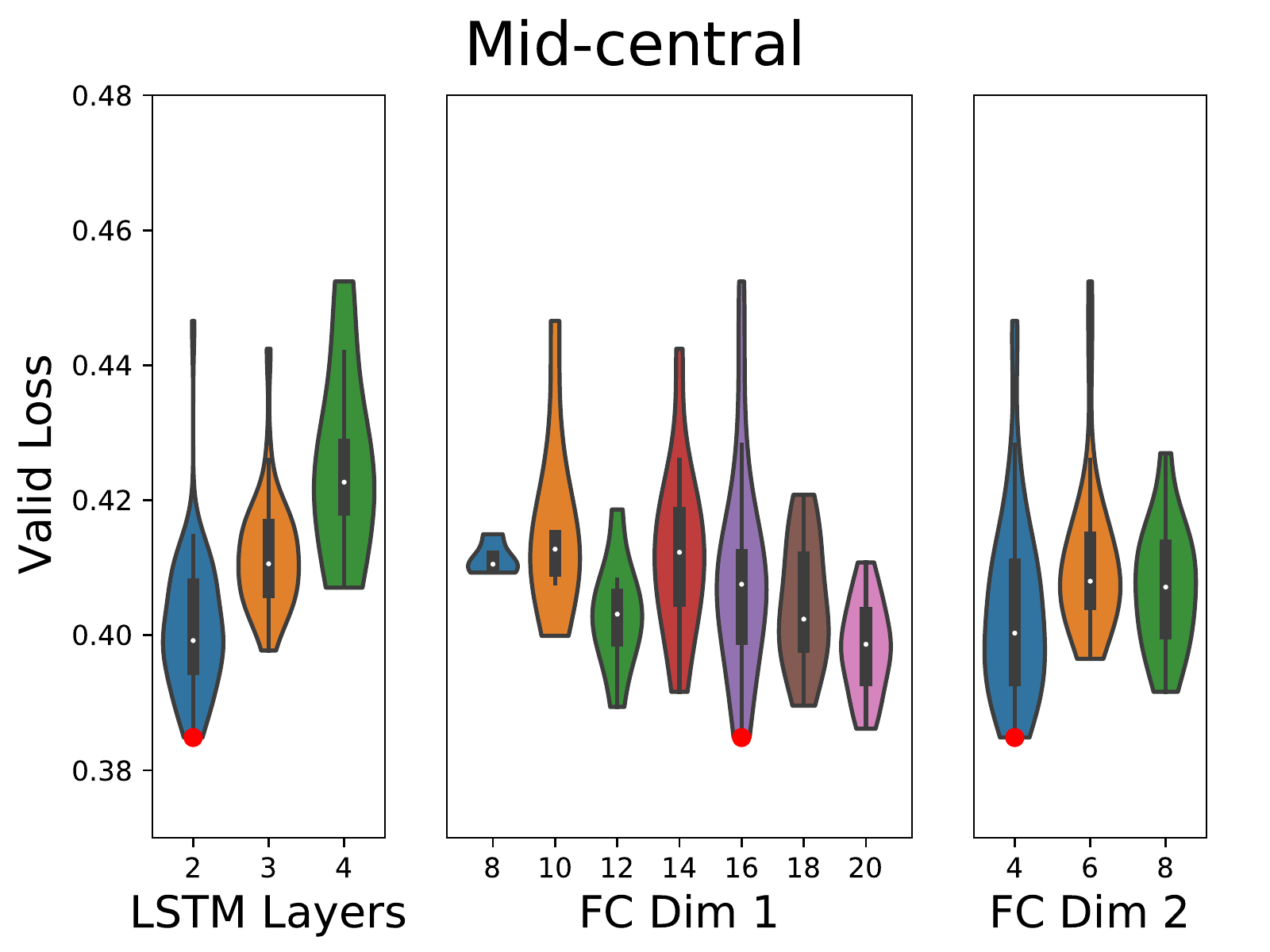}
    \includegraphics[width=0.49\textwidth]{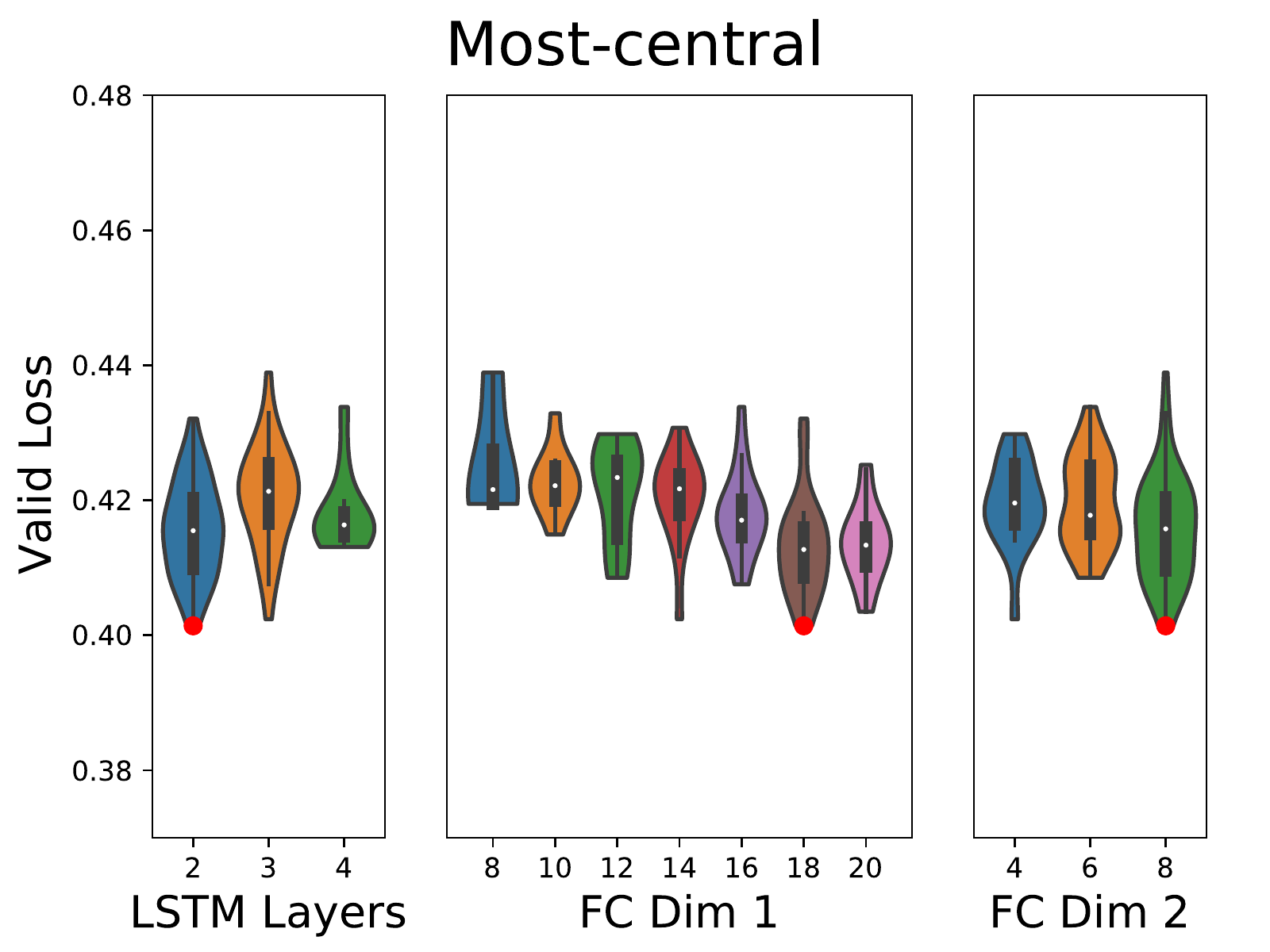}
    \caption{Distribution of validation loss for different architecture configurations, for the mid-central event mixing (left) and the most-central event mixing (right) respectively. The best models are indicated in red dots. The outer probability density contours cut off at the  min/max values of the validation loss. The inner white point represents the median value. The inner thick bar  spans the first and the third quartiles of the distributions.} 
    \label{fig:violin}
\end{figure}

The so-called hyper-parameters of a model, those that are not trainable but specify the details of the architecture and its training, are tuned with the Hyperopt~\cite{pmlr-v28-bergstra13} package. The hyper-tuning procedure helps locate the optimal hyper-parameter set on its search space, such that a local minimum can be avoided. The hyper-parameters related to the neural network architecture are the number of LSTM layers, and the input dimensions of two FC layers\footnote{The input dimension of the first FC layer is equal to the LSTM hidden size.}. The hyper-parameters related to the training process include the number of training epochs, batch size, learning rate and its decay factor. We scan over the hyper-parameter search space 50 times and arrange three trainings for each hyper-parameter set. The distribution of validation loss (defined in the same way as the batch loss, but calculated with the validation dataset) for discrete hyper-parameter options are displayed in Fig.~\ref{fig:violin} with violin plots, showing both the probability density of the validation loss and its quartiles. The trained model with minimal validation loss shown with the red dot is selected as the best trained model, with the optimal hyper-parameter values listed in Table~\ref{tab:tab2}.

\begin{table}[ht]
\begin{center}
\begin{tabular}{ c|c|c } 
 Parameters & Mid-central & Most-central\\ 
 \hline
 No. of LSTM Layers & 2 & 2\\ 
 FC Dim 1 & 16 & 18\\
 FC Dim 2 & 4 & 8 \\
 \hline
 No. of Epochs & 50 & 45\\
 Batch Size& 10000 & 18000\\
 Learning Rate & 0.0250 & 0.0371\\
 Decay Factor & 0.9689 & 0.9896\\
 \hline
\end{tabular}
\bigskip
\caption{List of hyper-parameters related to the neural network architecture and the training process, and their optimal values after hyper-tuning.}
\label{tab:tab2}
\end{center}
\end{table}

\subsection{Robustness}
\begin{figure}[htp]
    \centering
    
    \includegraphics[width=0.328\textwidth]{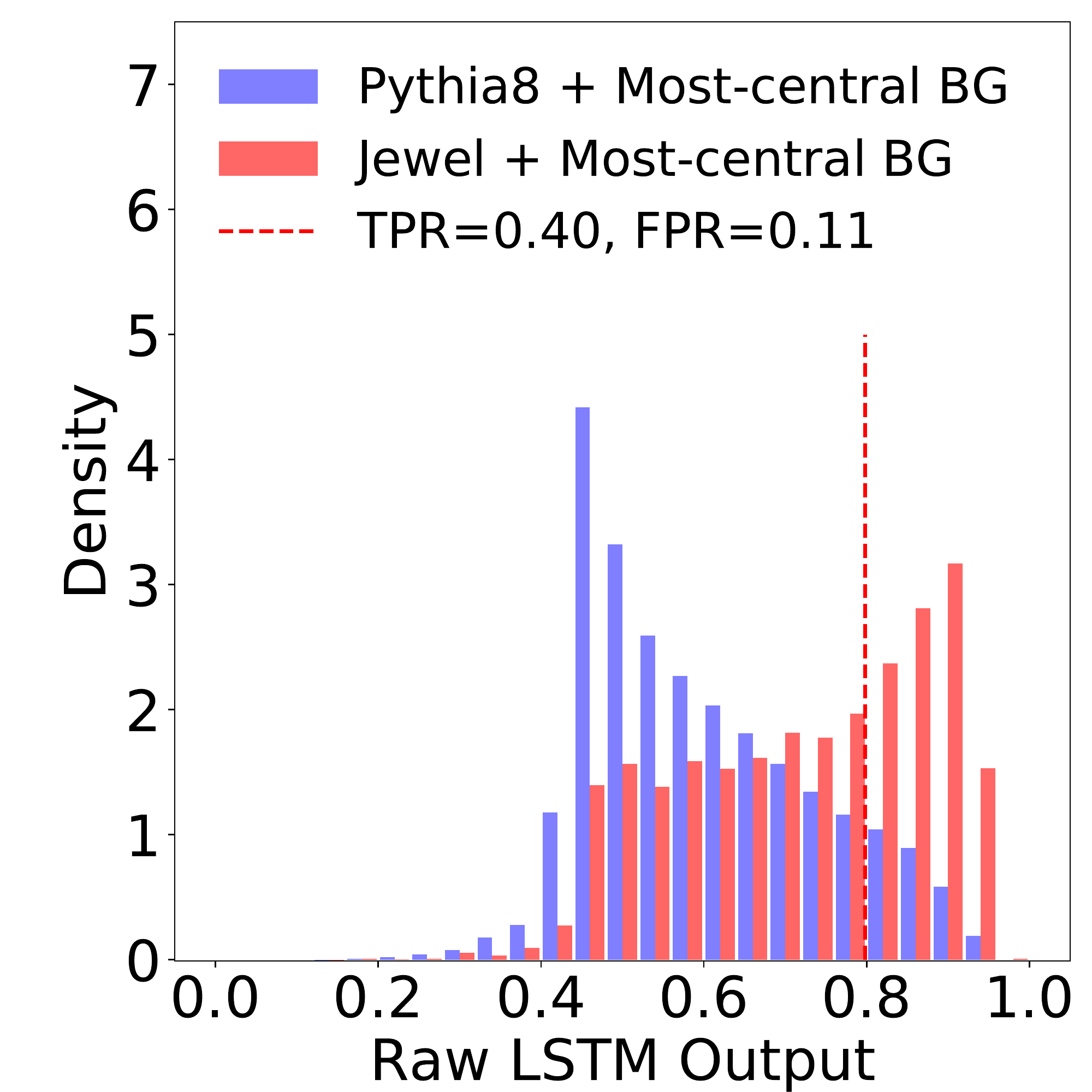}
    \includegraphics[width=0.328\textwidth]{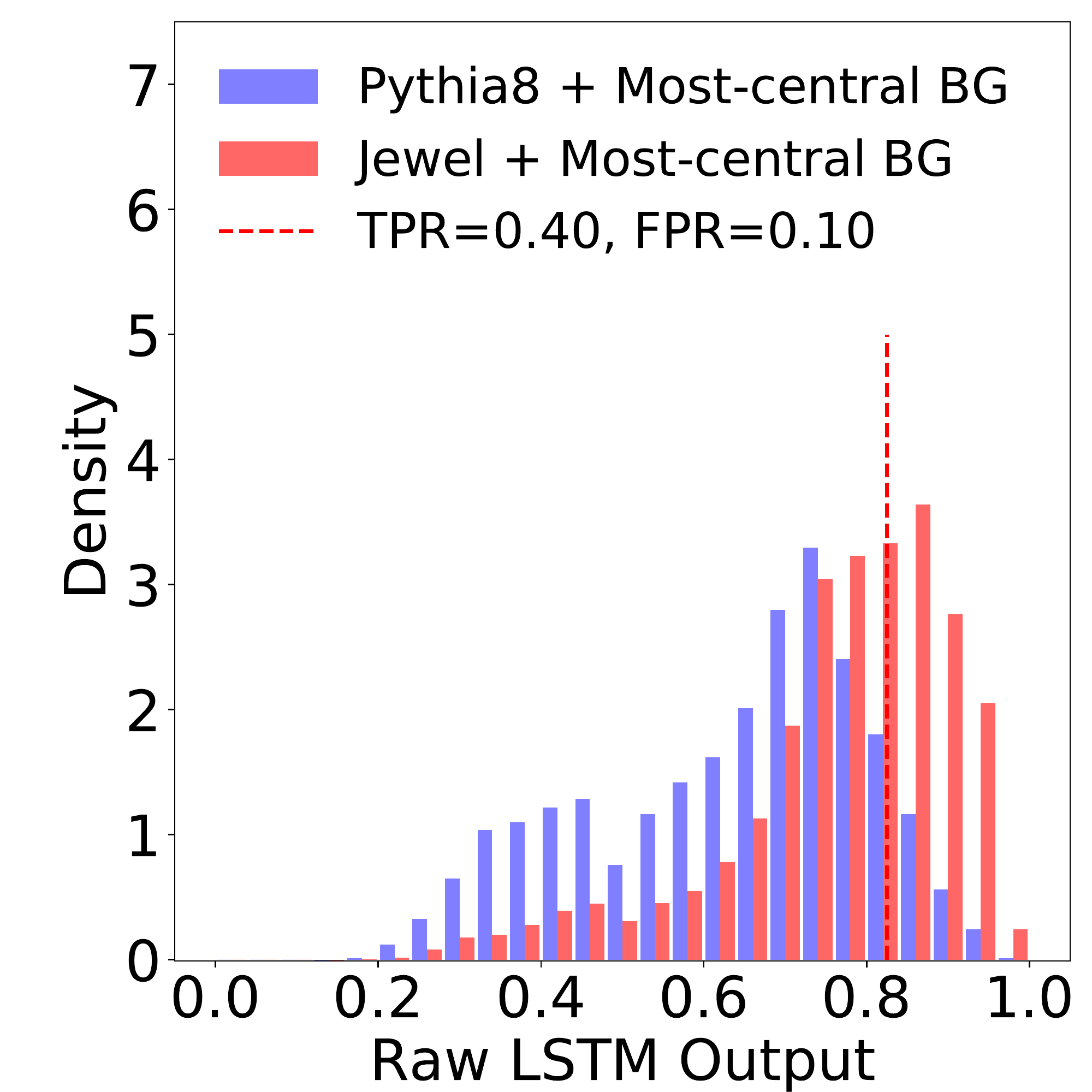}
    \includegraphics[width=0.328\textwidth]{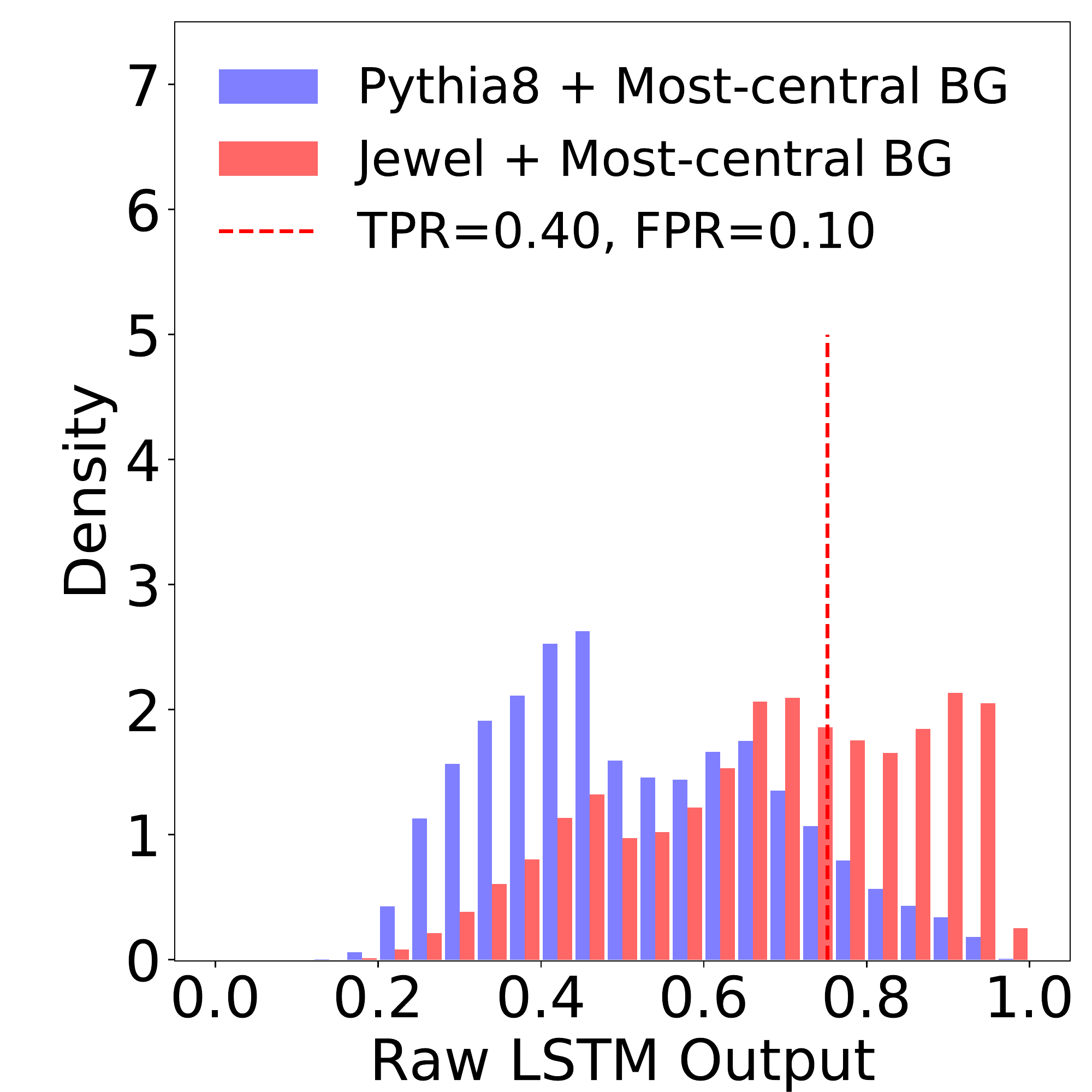}
    
    \vspace{0.1in}
    
    \includegraphics[width=0.328\textwidth]{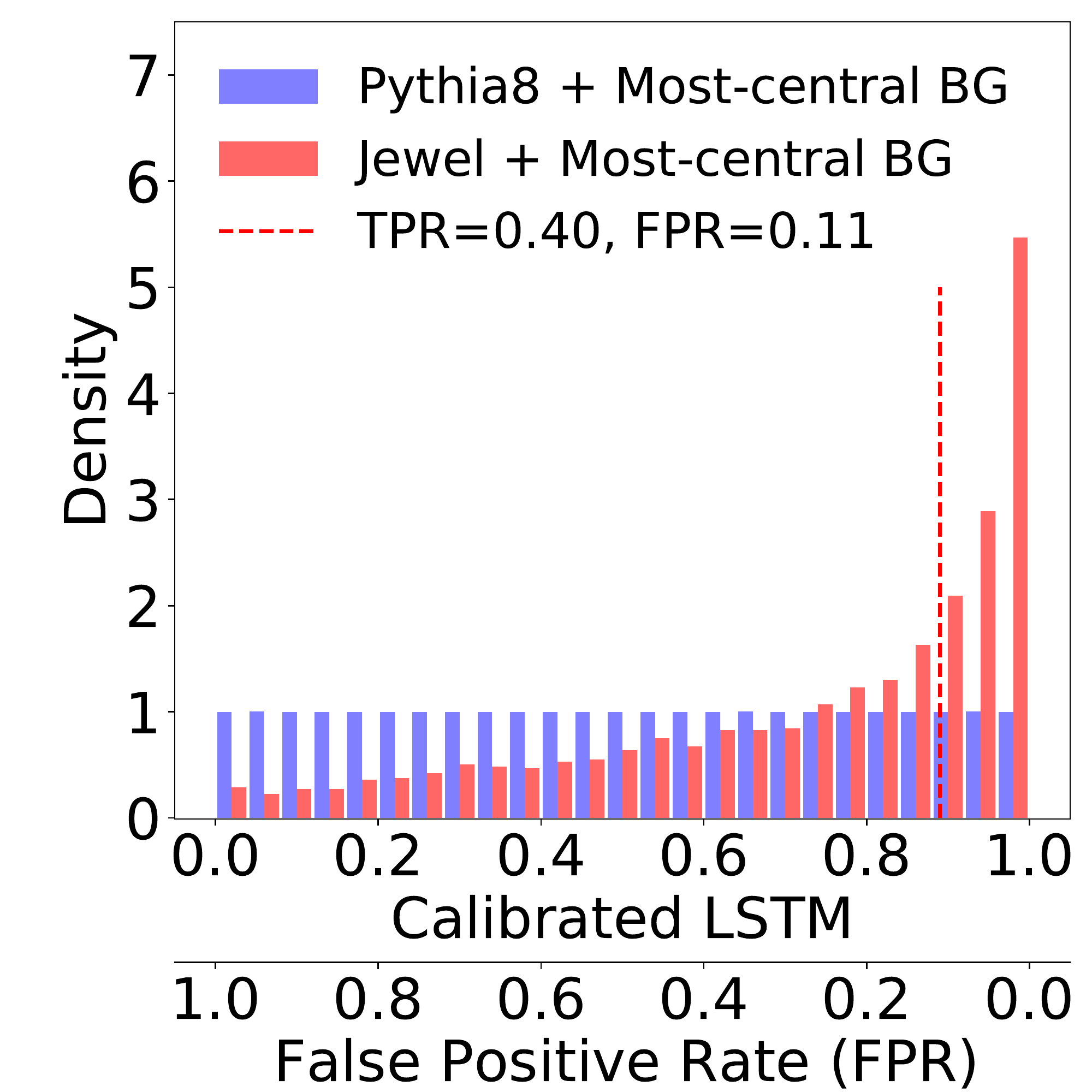}
    \includegraphics[width=0.328\textwidth]{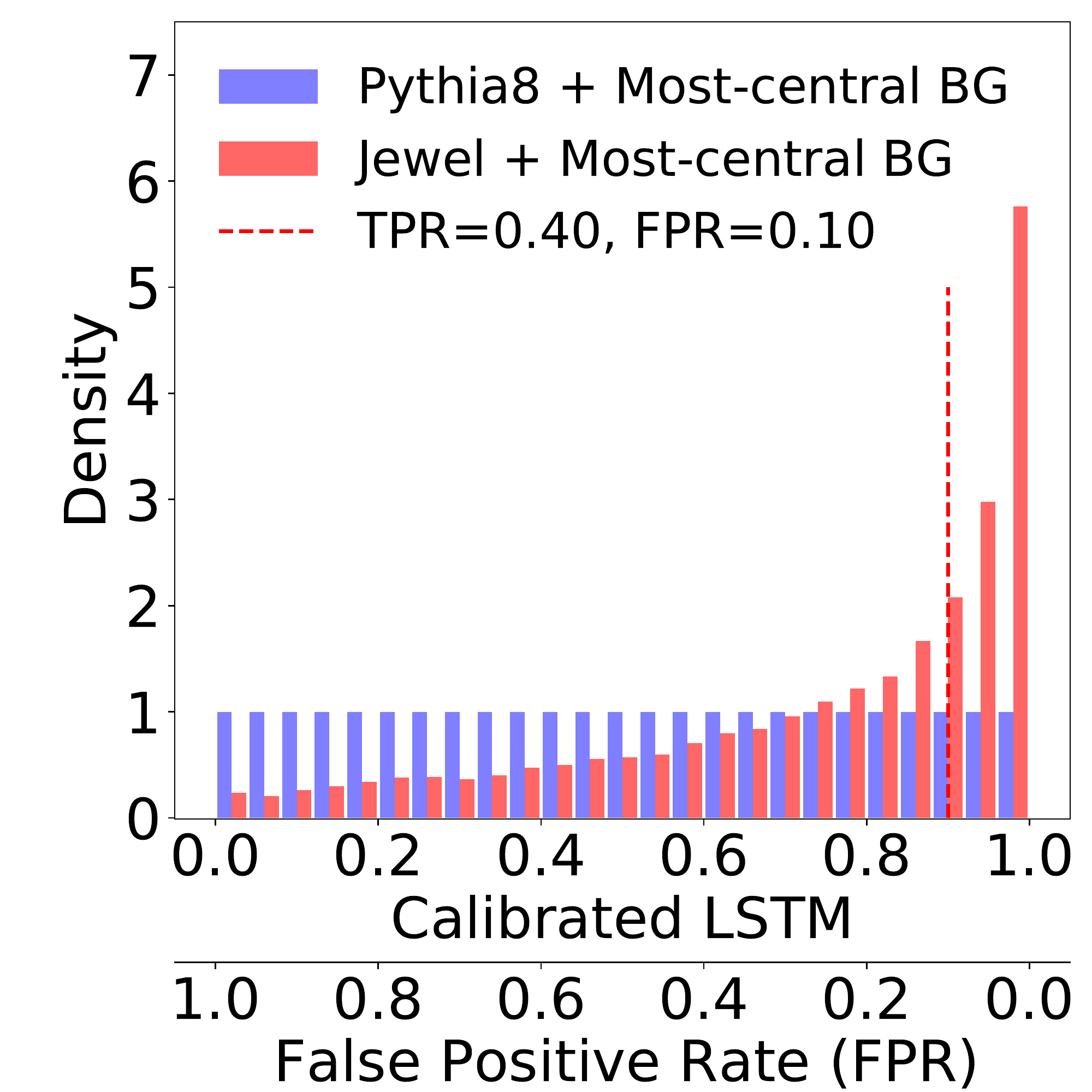}
    \includegraphics[width=0.328\textwidth]{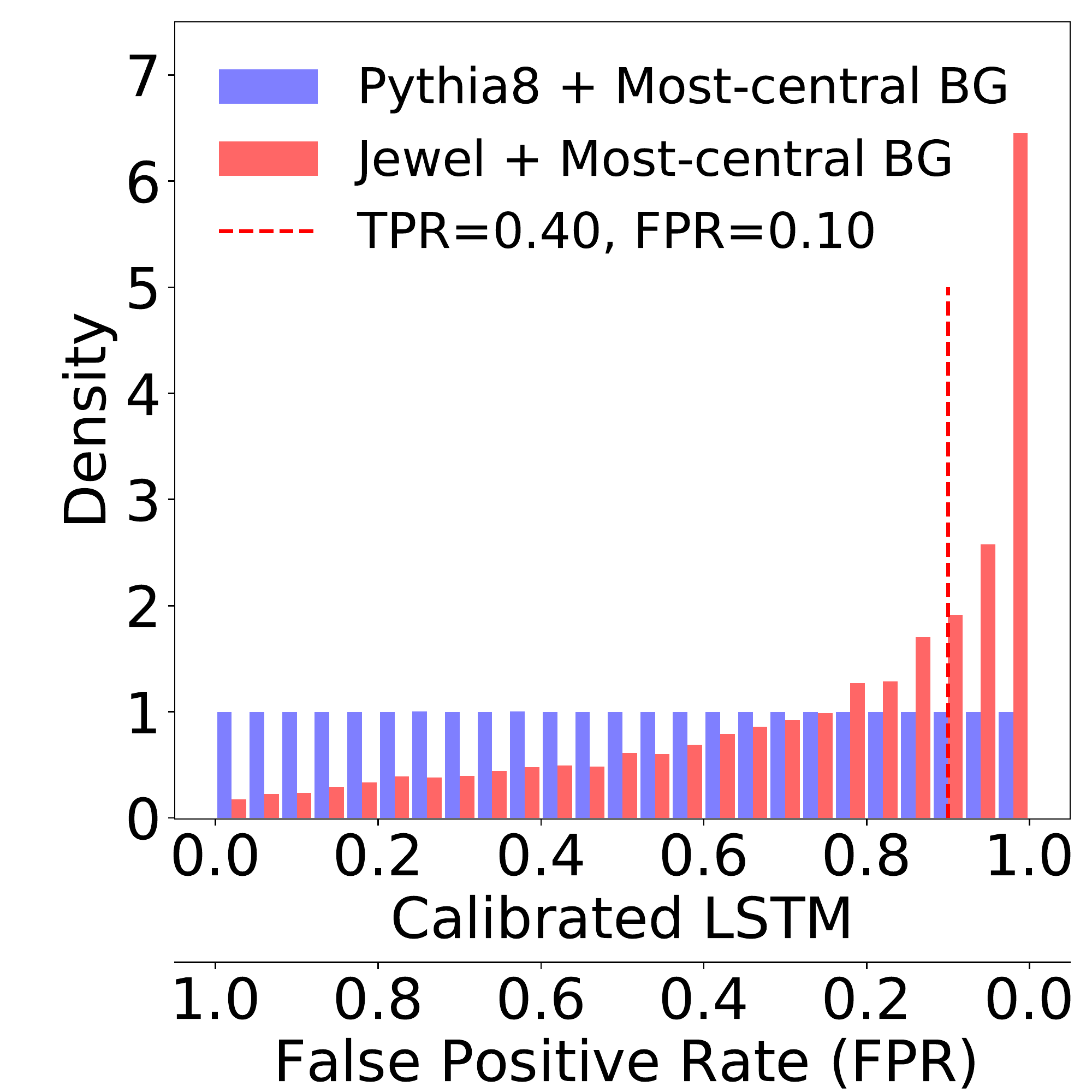}
    
    \vspace{0.1in}
    
    \includegraphics[width=0.328\textwidth]{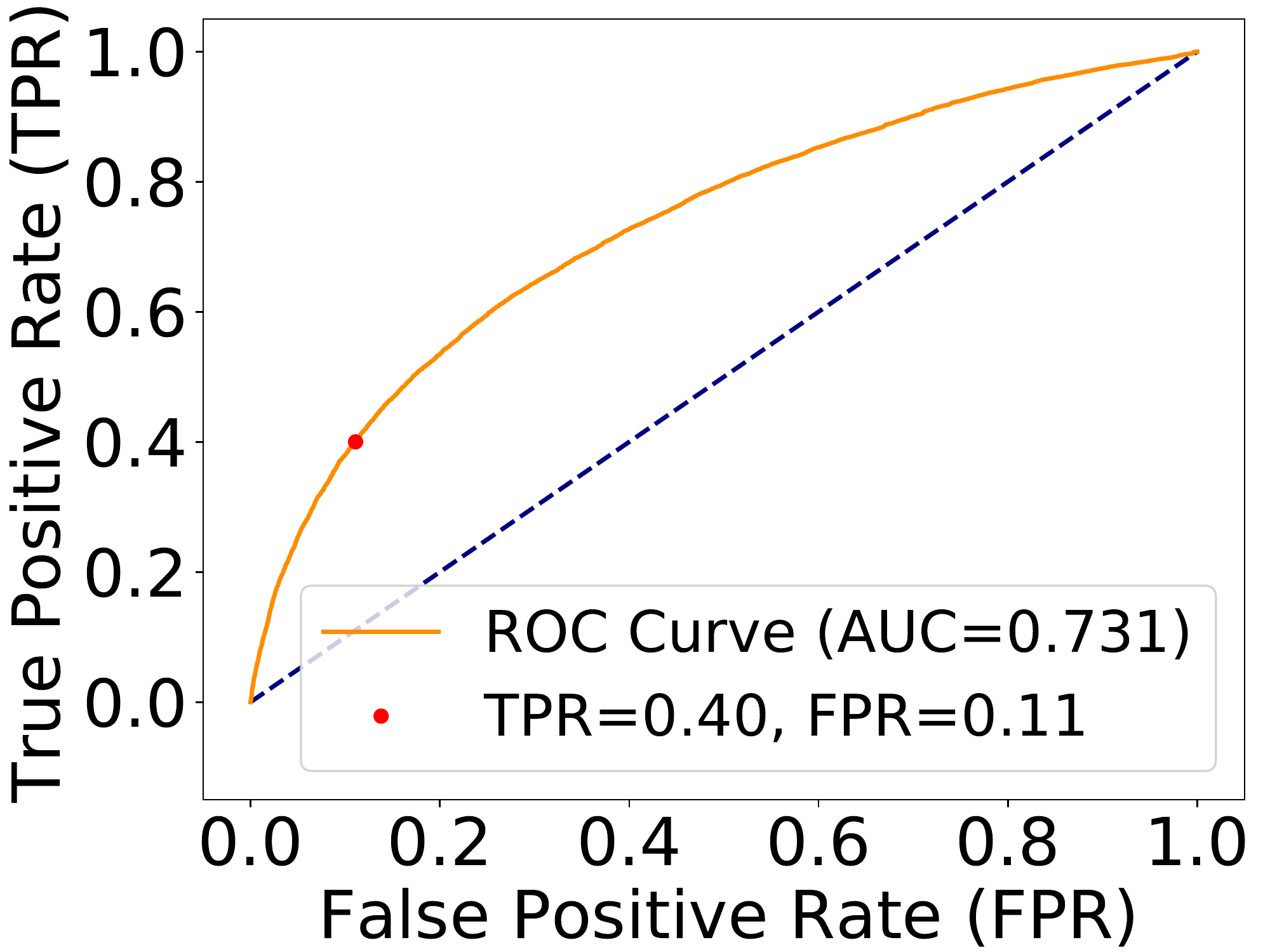}
    \includegraphics[width=0.328\textwidth]{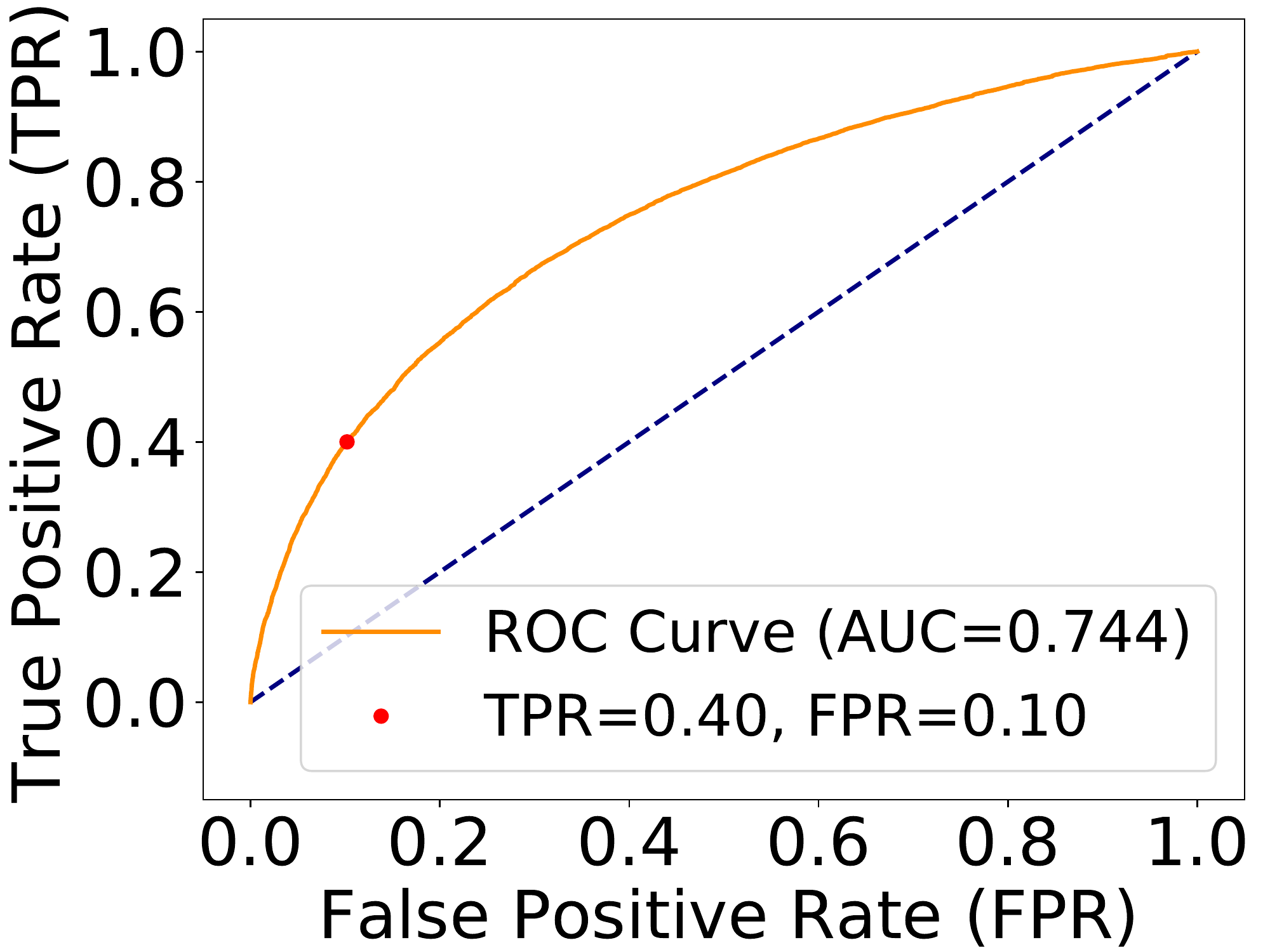}
    \includegraphics[width=0.328\textwidth]{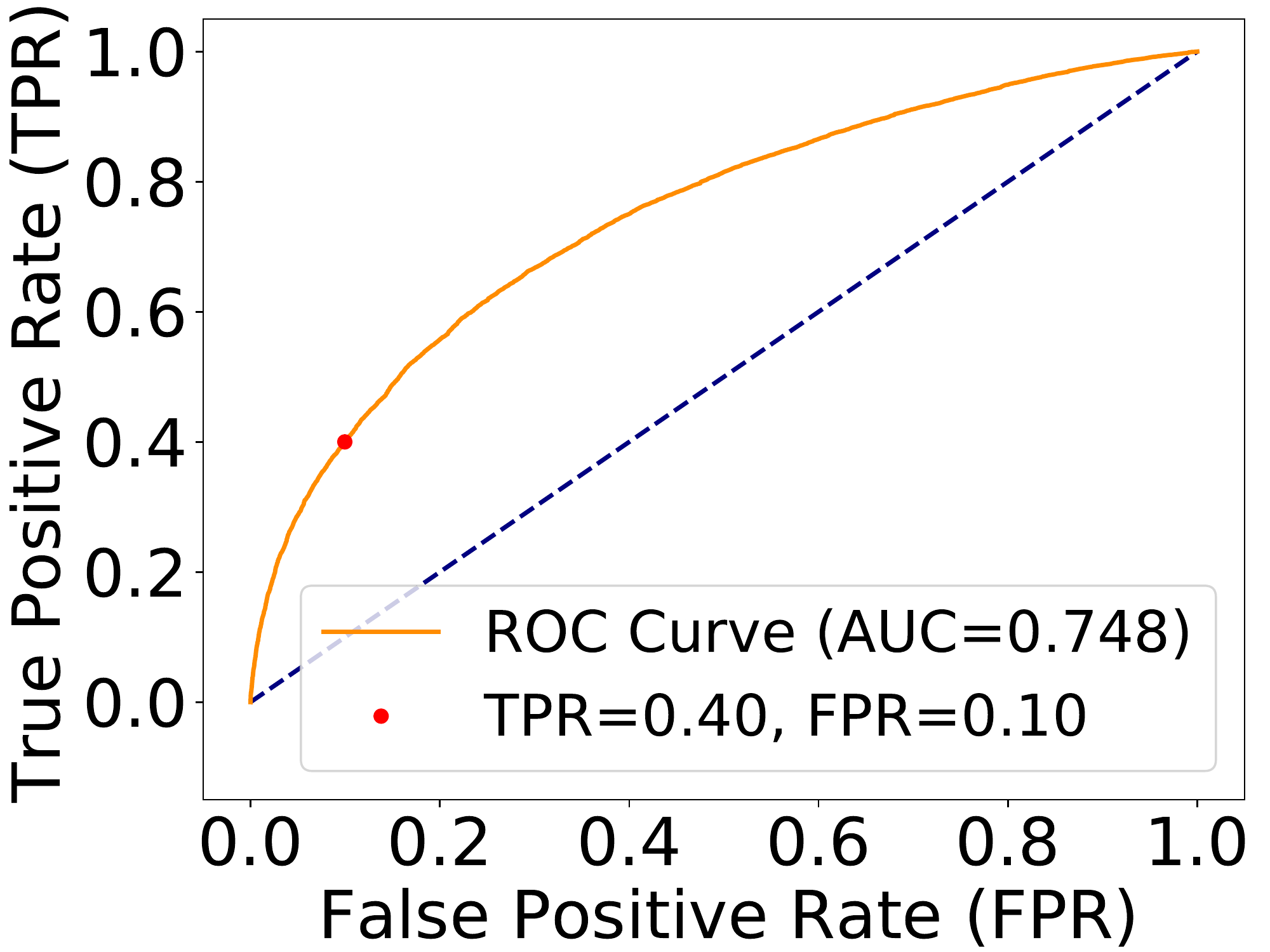}
    
    \vspace{0.1in}
    
    \caption{Distributions of the raw LSTM output from top three best-trained neural networks (top panels). Distribution of calibrated LSTM output (middle panels), and the corresponding ROC curve (bottom panels). The red lines/points corresponds to  thresholds determined with TPR=0.4.} 
    \label{fig:calib}
\end{figure}

The trained neural network can be used as a classifier, making predictions on the probability for a jet to be quenched. Here, the distributions of the raw LSTM output from the top three best-trained networks are shown in Fig.~\ref{fig:calib}, which exhibit a non-deterministic behavior in the metric. During the training, samples of the negative class are labeled with 0, while samples of the positive class are labeled with 1, without intermediate labels in between. Hence, the metric that quantifies the probability for a jet to be quenched, which is established during the training process, suffer from the stochastic nature of machine learning. Such non-determinism is rooted in the randomness of selecting samples to form a batch. To solve this issue, a calibration method is designed.

The calibration method is designed such that samples of the negative class are uniformly distributed along the newly established metric, meanwhile the relative quantitative relationships between the samples' raw LSTM outputs are kept in order. The new metric can be obtained with the help of the receiver operating curve (ROC), which plots the true positive rate (TPR) against the false positive rate (FPR) at various threshold settings. The FPR quantifies the percentage of samples in the negative class that are falsely predicted as positive. For illustration, with a decision threshold of 0, all negative samples should have predictions larger than 0, which leads to FPR=1. The new metric is bound to the FPR, meanwhile the threshold on the raw LSTM output corresponding to a specific FPR value can be determined from the ROC curve. The distribution of the raw LSTM output from trained neural networks and their values after calibration are shown in Fig.~\ref{fig:calib} (top and middle panels, respectively). For the three well-trained classifiers our calibration method produces almost identical distributions of the calibrated LSTM values. 

Given the distributions of the LSTM outputs, either raw or calibrated, it can be seen that not all samples in the positive class (jets from \textsc{Jewel}) are predicted to be quenched by the trained neural network. To determine if a jet is quenched, a threshold should be used, such that jets that have predictions larger than such a threshold would be classified as quenched, while those with predictions smaller than the threshold would be considered  non-quenched. In the subsequent discussion use the threshold determined from TPR=0.4. With this threshold,  the positive samples from \textsc{Jewel} are divided into top 40\% subset,  and bottom 60\% subset. It can be seen from the red point on the ROC curve that  this TPR corresponds to a FPR of roughly 10\%, indicating that about 10\% of the samples in the negative class have predictions larger than the threshold. Further, we consider the top 40\% \textsc{Jewel} jets as quenched and compare their substructure variables with the remaining 60\%, as well as those simulated with \textsc{Pythia}.


\section{Results}
\label{sec:results}

\begin{figure}[htpb]
    \centering
    \includegraphics[width=0.328\textwidth]{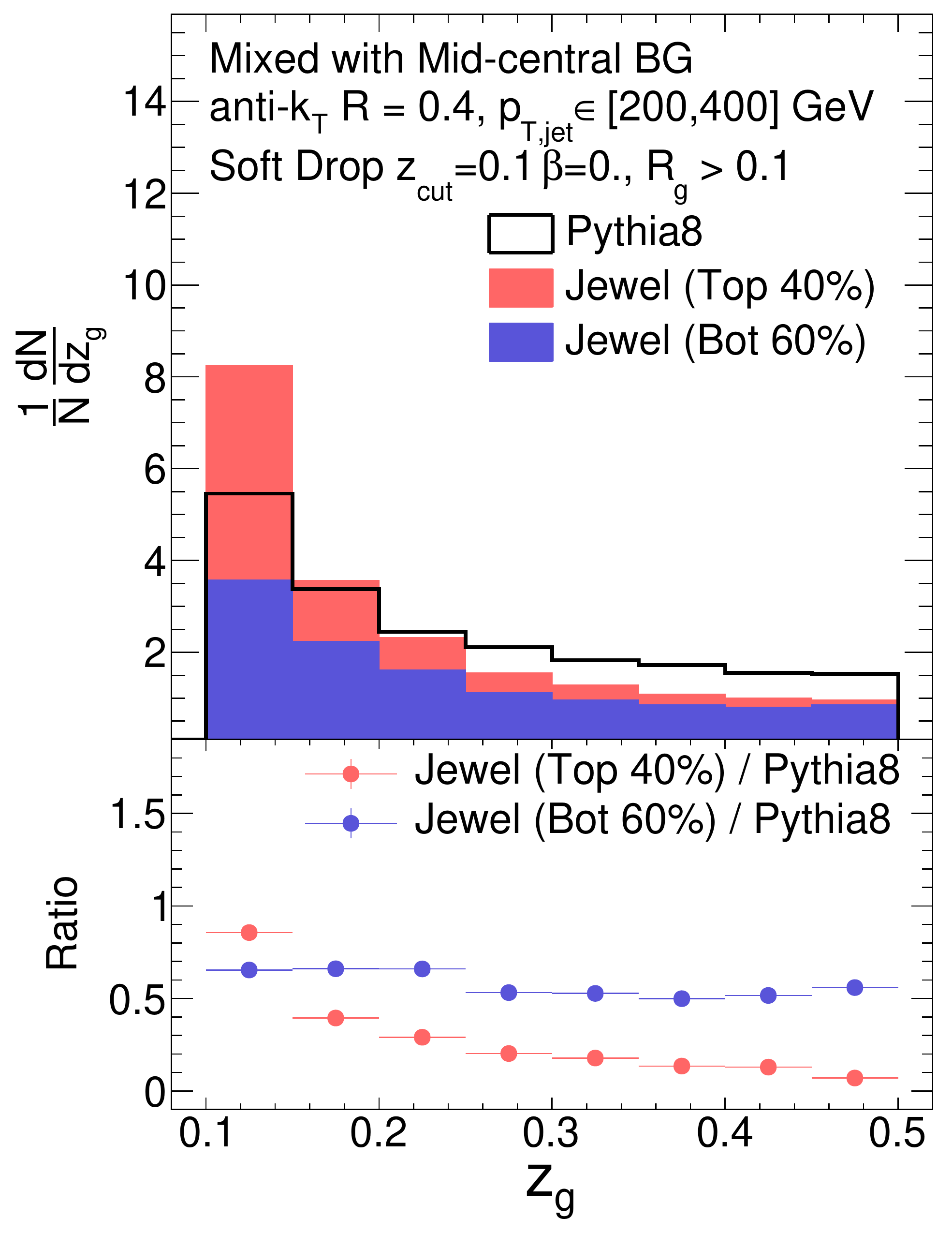}
    \includegraphics[width=0.328\textwidth]{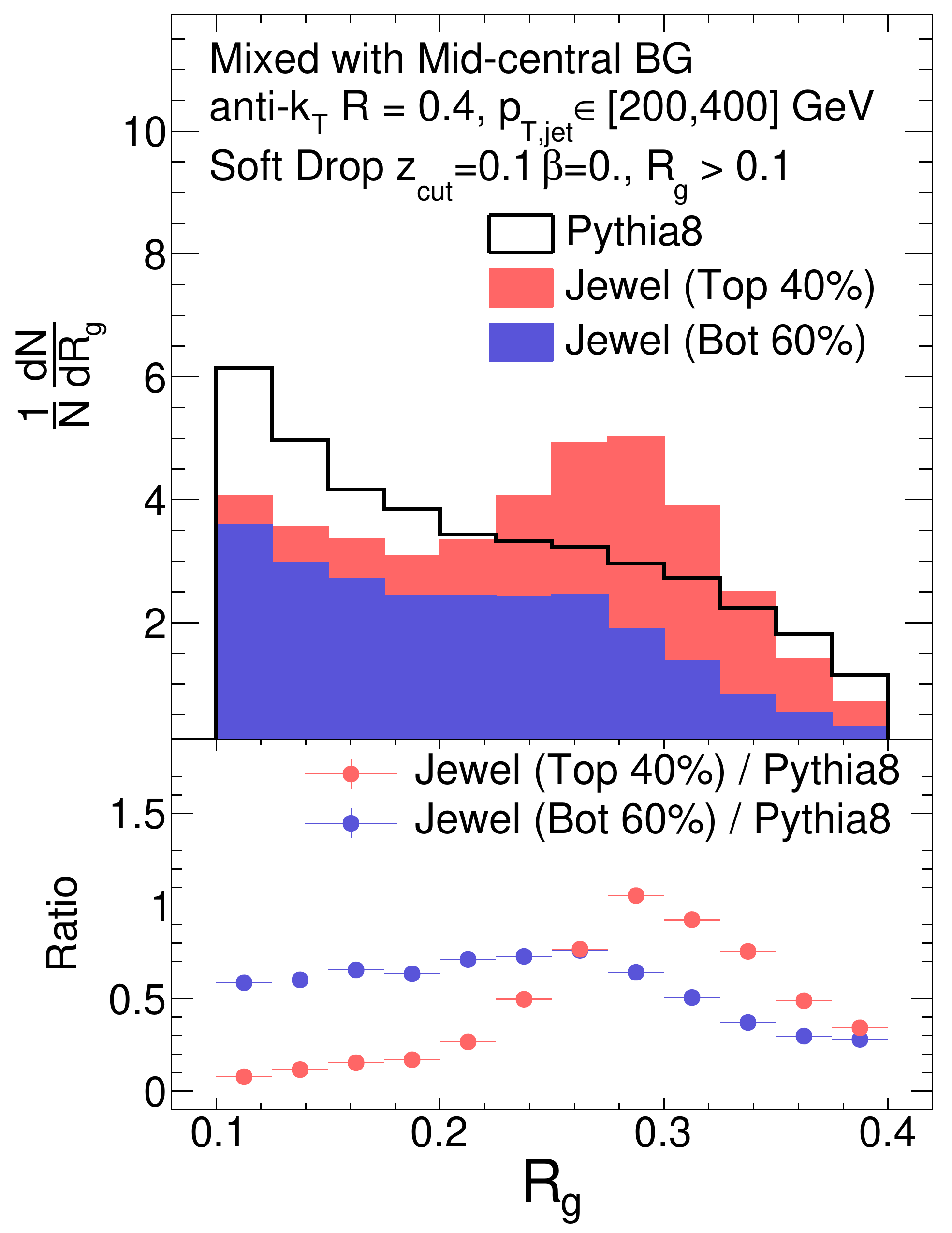}
    \includegraphics[width=0.328\textwidth]{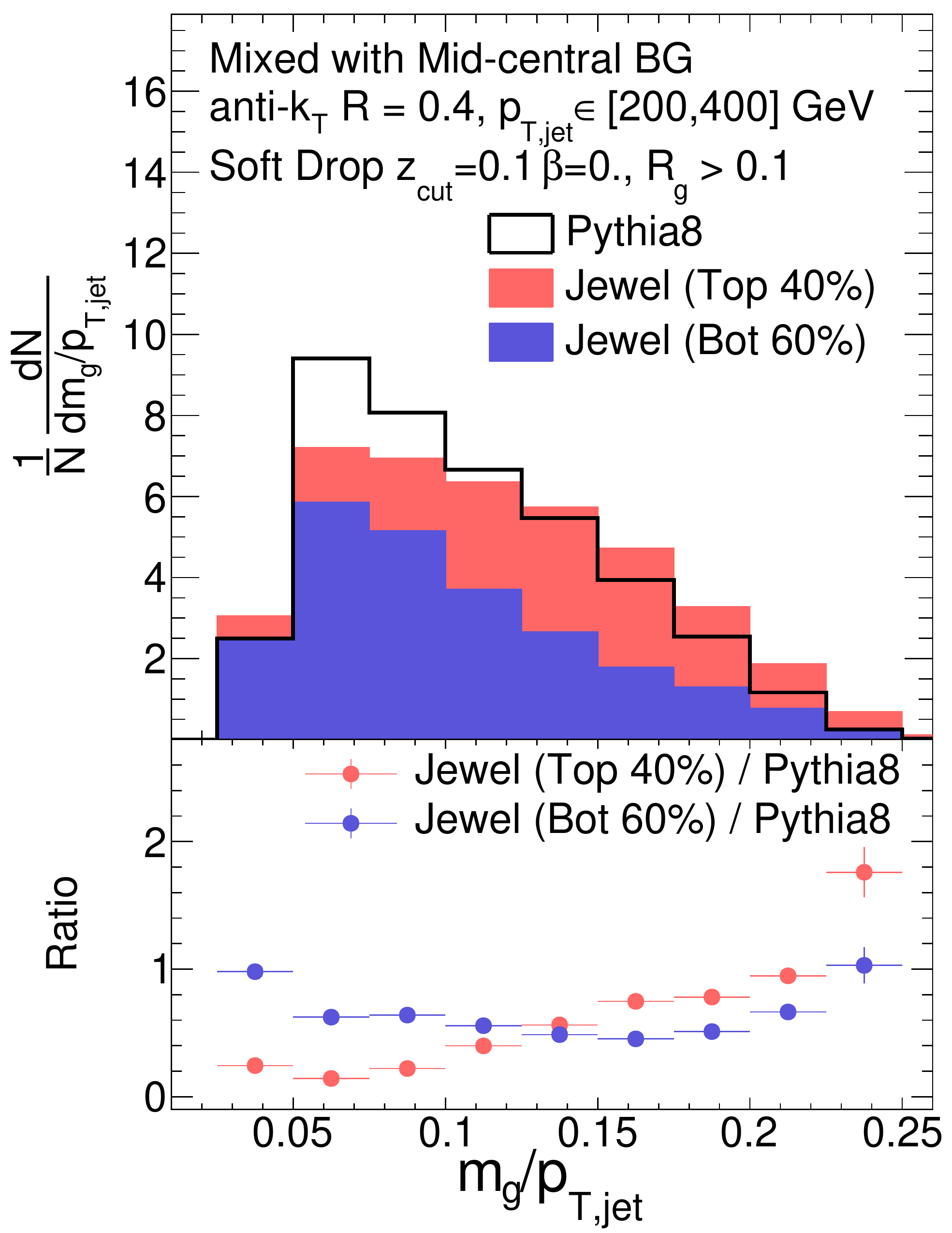}
    
    \vspace{0.1in}
    
    \includegraphics[width=0.328\textwidth]{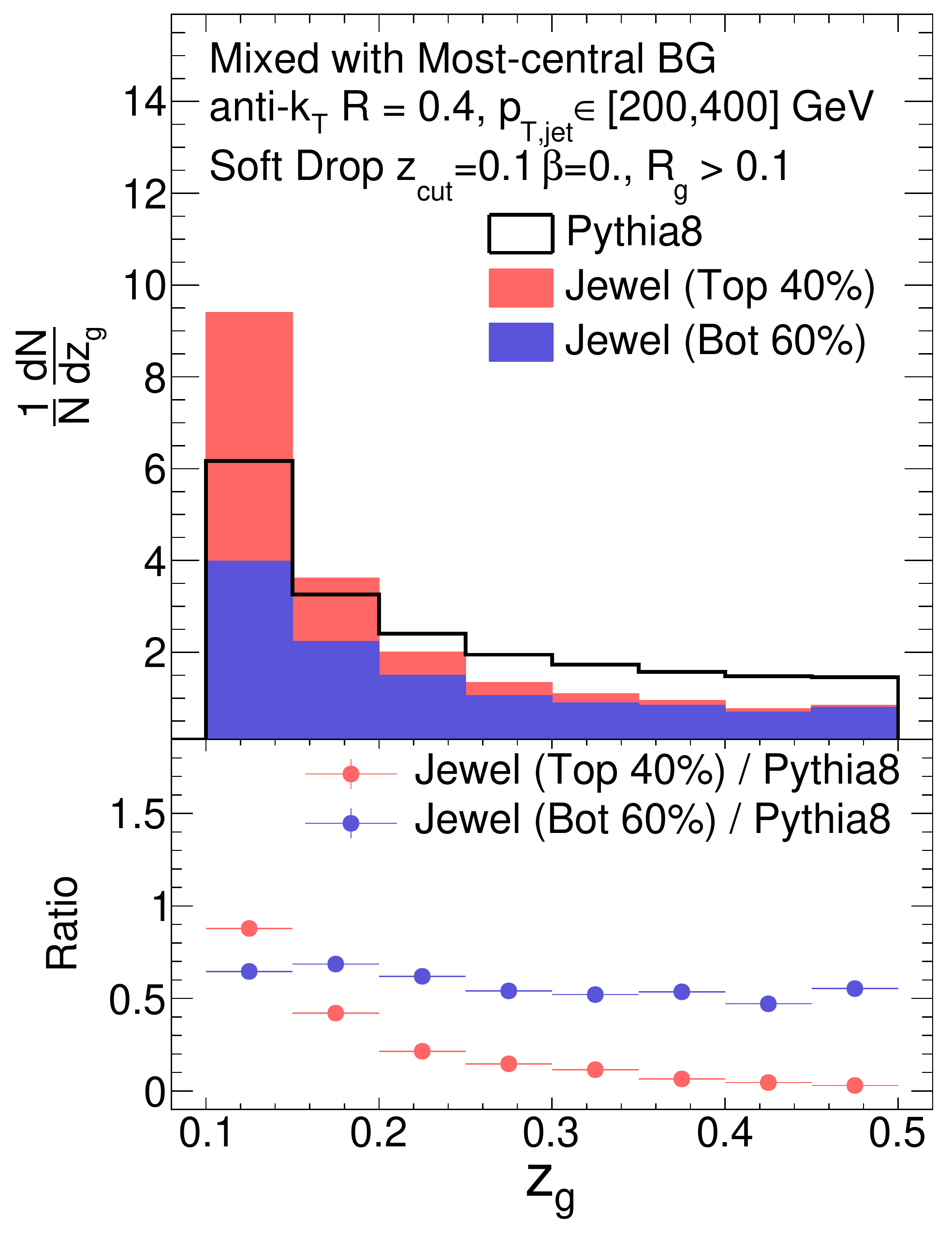}
    \includegraphics[width=0.328\textwidth]{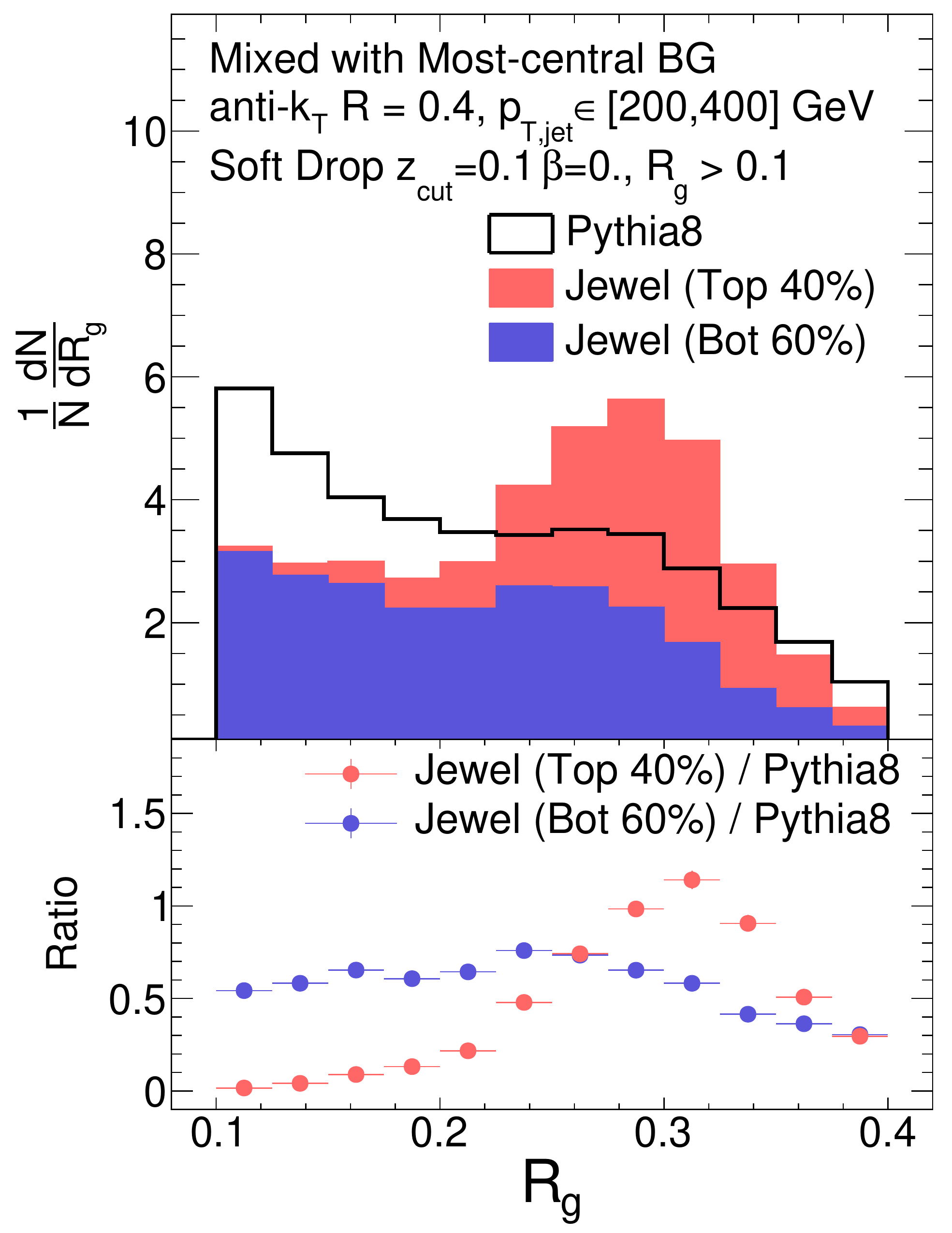}
    \includegraphics[width=0.328\textwidth]{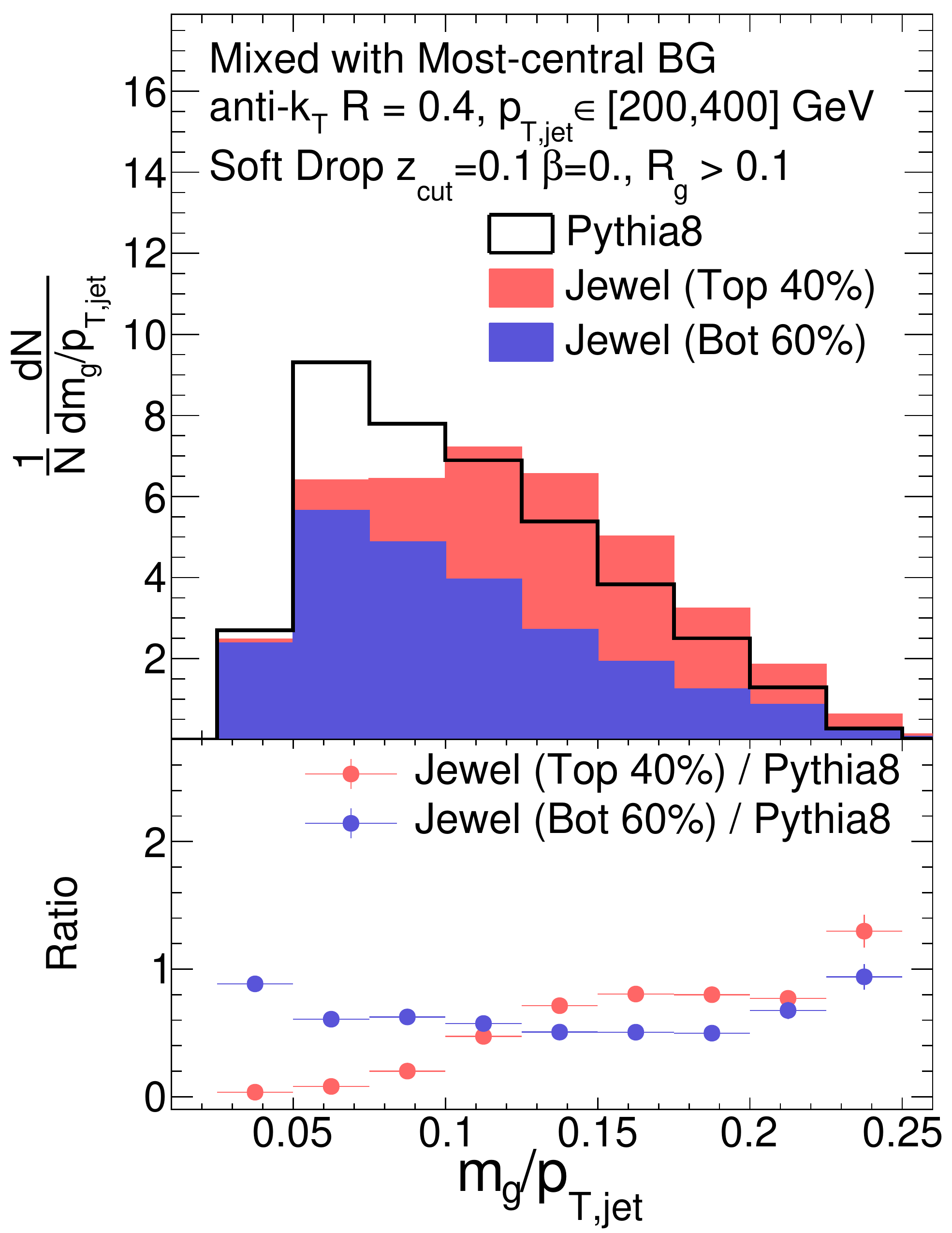}
    
    \caption{Distribution of the substructure variables of the groomed jets in the mid-central events (top) and the most-central events (bottom). Samples of the positive class (\textsc{Jewel}) are plotted with stacked histograms.} 
    \label{fig:stack}
\end{figure}

\begin{figure}[htpb]
    \centering
    \includegraphics[width=0.48\textwidth]{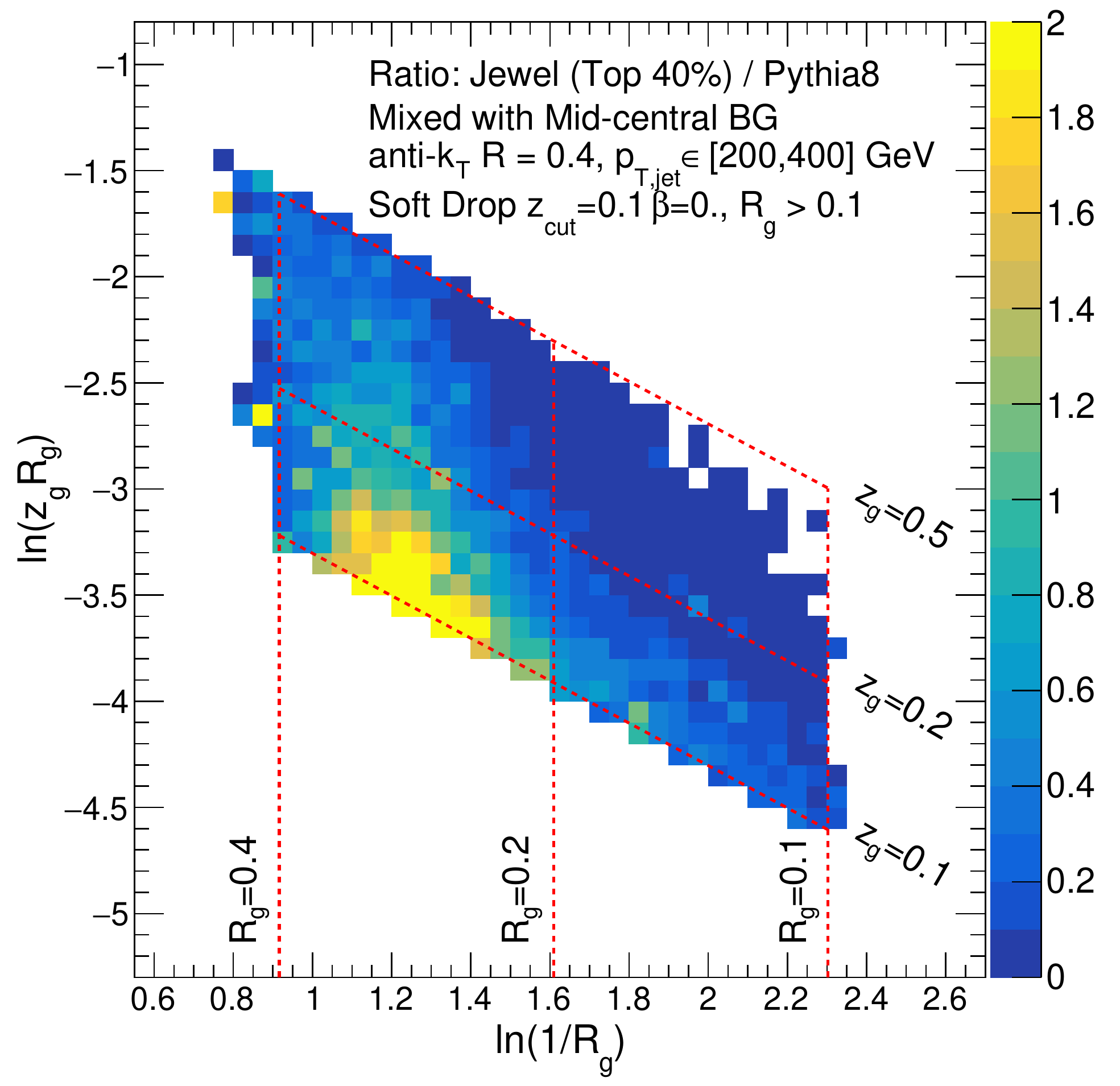}
    \includegraphics[width=0.48\textwidth]{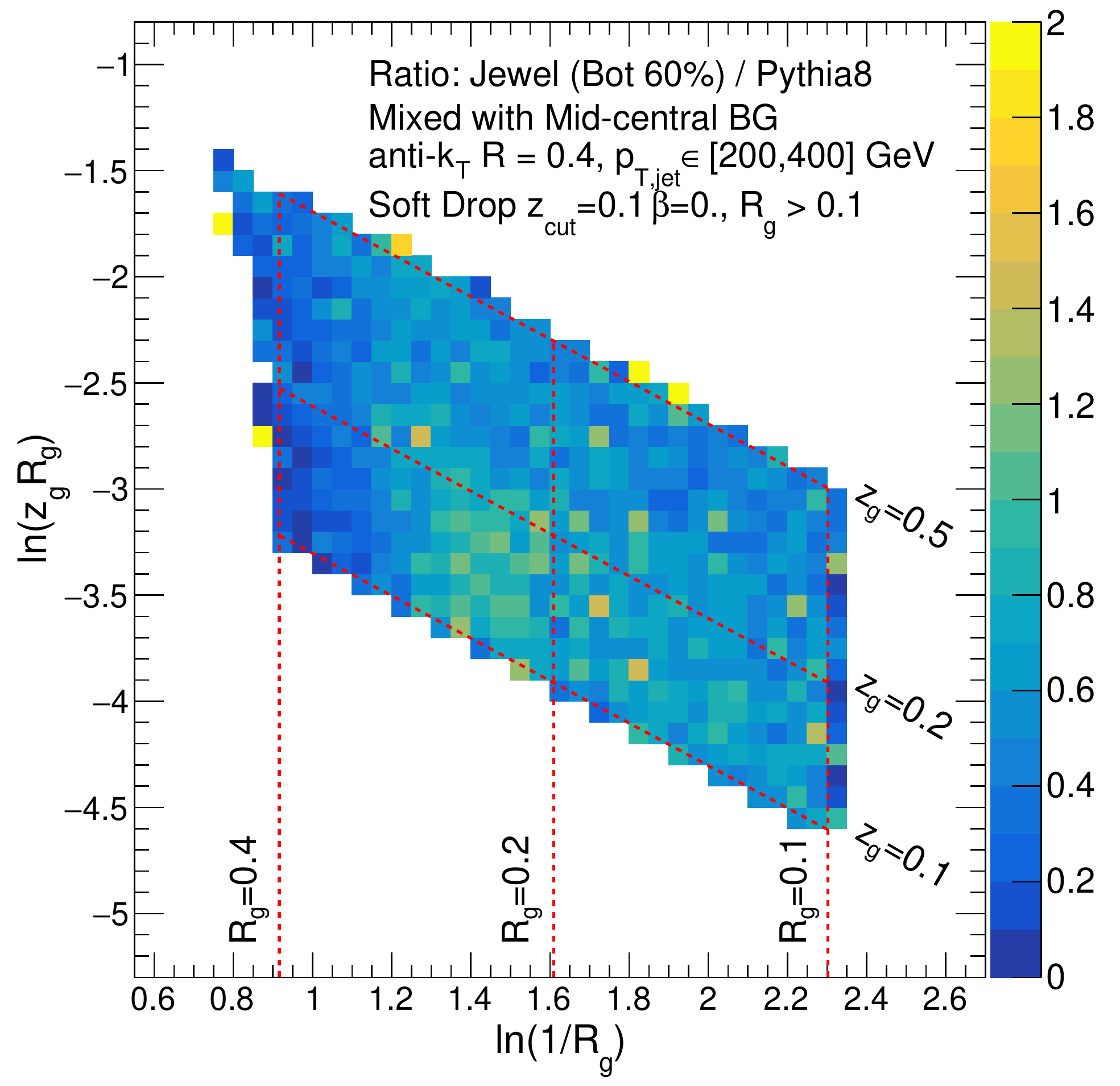}
    
    \vspace{0.1in}
    
    \includegraphics[width=0.48\textwidth]{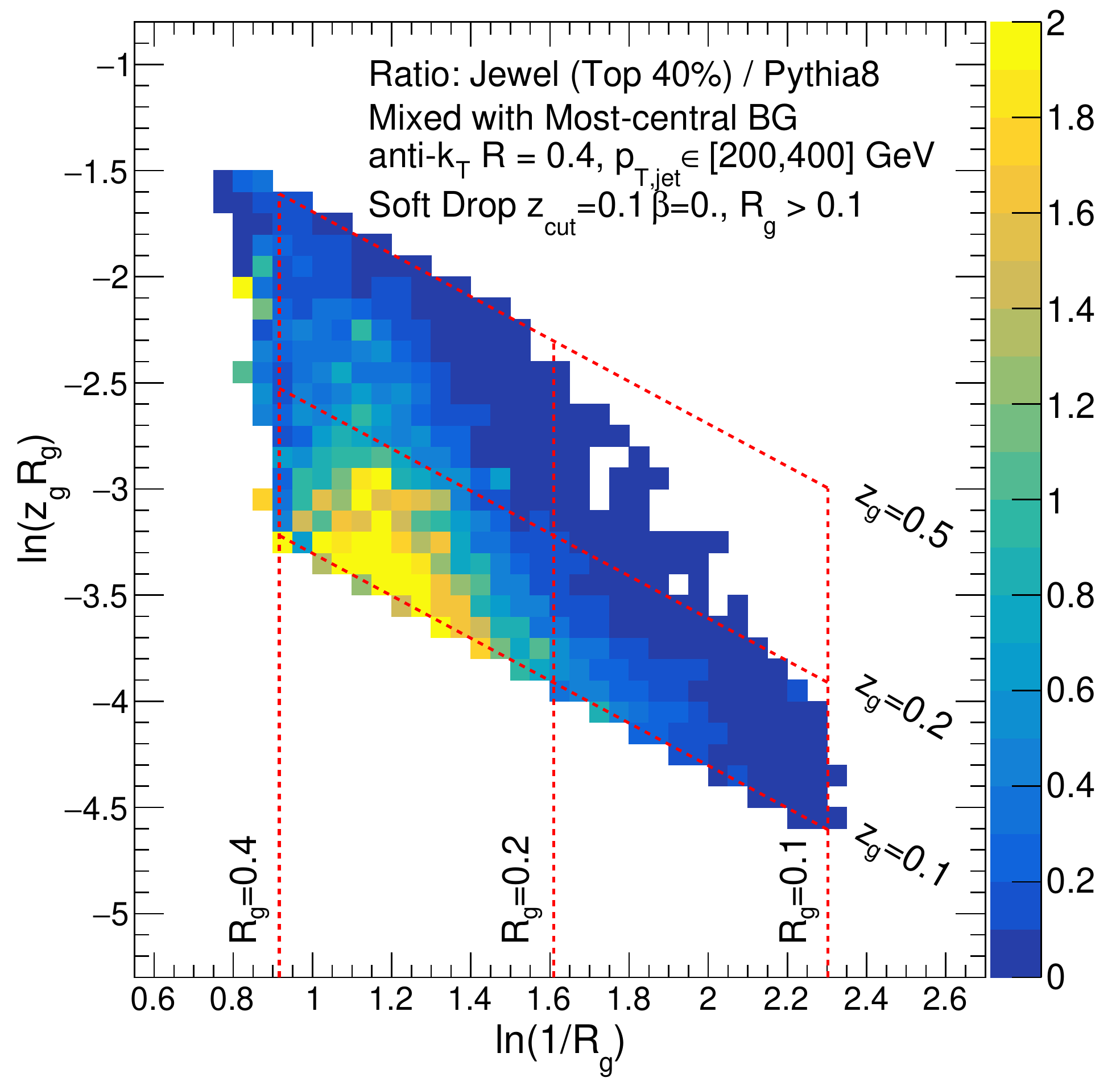}
    \includegraphics[width=0.48\textwidth]{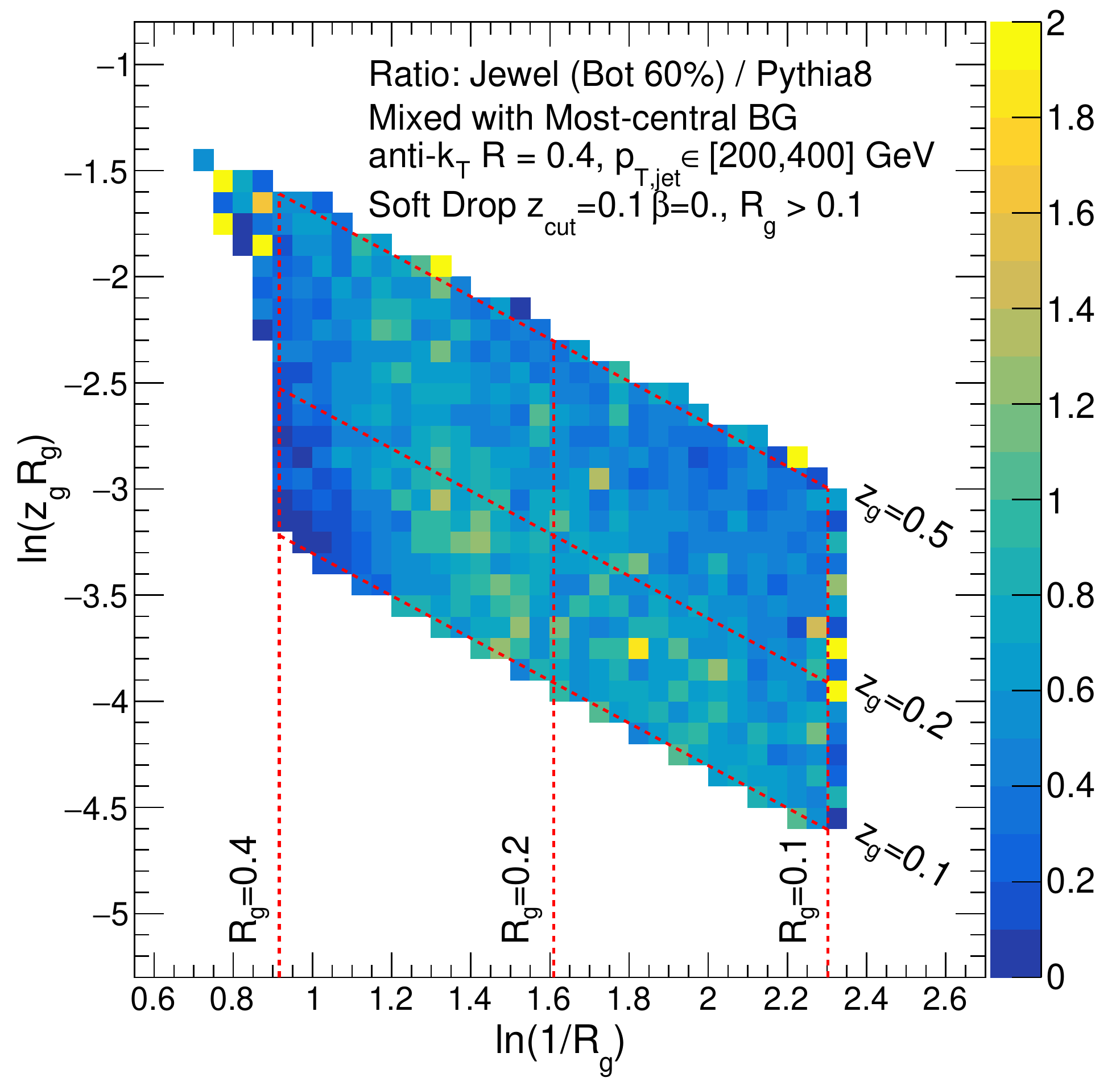}
    
    \caption{Primary Lund plane  density for the top 40\% \textsc{Jewel} samples (left) and the bottom 60\% \textsc{Jewel} samples (right). Plots for the mid-central and the most-central event mixing scenarios are shown in the upper and lower panels, respectively.} 
    \label{fig:lund_lstm}
\end{figure}

Samples of the positive class (\textsc{Jewel}) are divided into two subsets, consisting of top 40\% and bottom 60\% samples, respectively, based on the raw LSTM output. To better understand the relation between the LSTM output and the quenching effects, we plot the distributions of the substructure variables of the groomed jets in Fig.~\ref{fig:stack},  and the primary Lund plane density for the hard splittings after grooming in Fig.~\ref{fig:lund_lstm}, for two subsets separately. Samples of the negative class (\textsc{Pythia}) that serve as a baseline are also shown for comparison.

These two subsets of samples show quenching effects at different levels. The top 40\% samples of the positive class show severe modifications of the jet substructure with enhancement of wider and softer splittings. In comparison, the bottom 60\% samples exhibit a quenching pattern that is similar to that of the negative class. Two neural networks that are trained at two different multiplicity scenarios behave consistently, and show a similar ability in quantifying quenching effects. This indicates that the machine learning approach presented here successfully handles the uncorrelated underlying events independent of the background density. 


\section{Conclusion}
\label{sec:conclusion}

Our study has shown a promising machine learning approach to identifying jet quenching in the presence of uncorrelated underlying event background. Sequential substructure variables extracted from the jet clustering history carry information about the parton showering process with or without quenching effects, and can be learned by an LSTM neural network. The well-trained neural network can then be used to select jets at different quenching levels. We also studied the non-deterministic behavior of the supervised machine learning algorithm. With the help of the ROC curve, a calibration method is designed to avoid the non-determinism.

The procedure of extracting sequential substructure variables can be used as a general feature engineering method on jets. In comparison with jet image approaches, which utilize jet substructure in image pre-processing, our feature engineering method extracts sequential variables directly and explicitly from the jet substructure. The angular-ordered clustering history, which is in the basis of the sequential features carries information about the parton showering process. 

In the supervised machine learning setup, the truth labels, either quenched or non-quenched, are related to two different Monte Carlo simulators that produce two types of dijet events. In a real experiment, such labels can simply be replaced with the collision systems, either pp collisions or heavy ion collisions. Unlike the study in Ref.~\cite{Du:2020pmp} which uses the jet energy loss ratio $\chi_{\mathrm{jh}}$ in a regression setup, out approach does not rely on simulation-only variables. 

The well-trained neural network is capable of quantifying quenching effects, making it possible to divide jets simulated from a quenching model (\textsc{Jewel}) into two subsets. The two separated subsets of jets have shown differences in the jet substructure variables, related to two different parton showering patterns: a vacuum-like pattern, similar to what is implemented in \textsc{Pythia 8}, and a medium-like pattern, which produces highly quenched jets. Moreover, the medium-like pattern tends to mark up a zone on the Lund diagram, which may help extract information about medium properties. As pointed in Ref.~\cite{Andrews:2018jcm}, the jet transport coefficient $\hat q$, which is of interest, is connected to boundaries of the quenching zone on the Lund diagram.

The techniques involved in our study, such as the background subtraction and the jet grooming, have been previously applied to experimental data and their effectiveness is well-studied. Thus, our machine learning approach is very promising in the exploration of different jet classification topics with real experimental data.


\acknowledgments
We thank James Mulligan, Raghav Kunnawalkam Elayavalli, and Yi Chen for their helpful suggestions. We thank the Advanced Computing Center for Research and Education (ACCRE) at Vanderbilt University for providing computing resources. This work was supported in part by the US Department of Energy Grant No. DE-FG05-92ER40712, and by the Alfred P. Sloan Foundation, whose contributions are gratefully acknowledged. 

\bibliographystyle{JHEP}
\bibliography{ref}

\end{document}